\documentclass[%
 aip,
 amsmath,amssymb,
 reprint,%
]{revtex4-2}

\usepackage{graphicx}
\usepackage{dcolumn}
\usepackage{bm}

\usepackage[utf8]{inputenc}
\usepackage[T1]{fontenc}
\usepackage{mathptmx}
\usepackage{etoolbox}
\usepackage{color, soul}

\makeatletter
\def\@email#1#2{%
 \endgroup
 \patchcmd{\titleblock@produce}
  {\frontmatter@RRAPformat}
  {\frontmatter@RRAPformat{\produce@RRAP{*#1\href{mailto:#2}{#2}}}\frontmatter@RRAPformat}
  {}{}
}%
\makeatother
\begin{document}

\preprint{AIP/123-QED}

\title{Device geometry dependent deterministic skyrmion generation from a skyrmionium}

\author{Adyashakti Dash}
 \affiliation {Laboratory for Nanomagnetism and Magnetic Materials, School of Physical Sciences, National Institute of Science Education and Research (NISER), HBNI, Jatni 752050, India}
 
\author{Brindaban Ojha}
\affiliation{Laboratory for Nanomagnetism and Magnetic Materials, School of Physical Sciences, National Institute of Science Education and Research (NISER), HBNI, Jatni 752050, India}

\author{Shaktiranjan Mohanty}
\affiliation{Laboratory for Nanomagnetism and Magnetic Materials, School of Physical Sciences, National Institute of Science Education and Research (NISER), HBNI, Jatni 752050, India}

\author{Ashish Kumar Moharana}
\affiliation{Laboratory for Nanomagnetism and Magnetic Materials, School of Physical Sciences, National Institute of Science Education and Research (NISER), HBNI, Jatni 752050, India}

\author{Subhankar Bedanta$^*$}
\affiliation{Laboratory for Nanomagnetism and Magnetic Materials, School of Physical Sciences, National Institute of Science Education and Research (NISER), HBNI, Jatni 752050, India}
\email{sbedanta@niser.ac.in}

\date{\today}

\begin{abstract}
A magnetic skyrmionium can be perceived as an association of two magnetic skyrmions with opposite topological charges. In this work, we have investigated the transformation of skyrmionium into multi-skyrmionic states via domain wall (DW) pairs in three different devices with variable geometric configurations. The same device geometries were considered for single ferromagnetic layer as well as synthetic antiferromagnetic (SAF) system. It is observed that by tuning the current density, deterministic generation of skyrmions is possible via the spin transfer torque (STT). The proposed device is efficiently adjustable to change the number of skyrmions . The results may lead to development of skyrmion-based devices for neuromorphic and unconventional computing.

\end{abstract}


\maketitle

Magnetic skyrmions are topologically protected, localized non-collinear textures which have antiparallel magnetization core with respect to the periphery.\cite{Skyrme4} Nanoscale size, gyrodynamics, and low driving current density make skyrmions promising for future spintronic applications such as information carriers \cite{chui2017magnetic}, microwave devices \cite{microwavedevice, microwave1}, spin-wave/magnon devices \cite{magnondevice, magnondevice1}, quantum computing \cite{Quantumcomputing, li2021magnetic}, logic devices \cite{logicdevice, logicdevice1} , digital information, data processing, and storage\cite{storage,storage1,storage2,storage3,storage4} etc. Skyrmions have been stabilized in a wide variety of magnetic structures like heavy metal (HM$_1$)/ferromagnet (FM)/HM$_2$ (or oxide) \cite{film1, jiang2017skyrmions,film3,film4jiang2015blowing,film5everschor2018perspective}. There are several underneath mechanisms to stabilize the skyrmions.\cite{heinze2011spontaneous, okubo2012multiple, montoya2017tailoring, tokura2020magnetic, garel1982phase, lin1973bubble, nagaosa2013topological} Among these,a fine tuning of interfacial Dzyaloshinskii-Moriya Interaction (DMI) and effective anisotropy energy can lead to stabilization of skyrmions in thin films.\cite{dzyaloshinsky1958thermodynamic, moriya1960anisotropic, ojha2021driving, liu2015static} Till now, several ways have been proposed for skyrmion motion via spin-polarized current \cite{spindeger2019current, spinlin2013driven, spinliu2015switching}, magnetic field\cite{magfielddynamics,magfieldbuttner2015dynamics}, anisotropic gradient\cite{anisotropygradPhysRev} and so on\cite{film5everschor2018perspective, fert2013skyrmions}.

However, the controlled motion of skyrmions is a big challenge for practical application due to the skyrmion Hall effect (SkHE) \cite{SkHE2012NatPh, SkHEjiang2017direct, SkHElitzius2017skyrmion}. To overcome this significant obstacle caused by the magnus force, acting on the moving skyrmion, few strategies have been proposed\cite{overcomeSkhezhang2017magnetic, toscano2020suppression}. A promising way to minimize SkHE is by constructing an antiferromagnetically exchange-coupled bi-layer system which is commonly known as synthetic antiferromagnet (SAF), in which a skyrmion in the top layer accounting for Q = -1 and a skyrmion in the bottom layer accounting for Q = +1 form a common entity with Q = 0. \cite{safzhang2016magnetic, safzhang2016thermally} Further, skyrmionium, a composite doughnut-like structure with a core skyrmion with upward magnetization and peripheral skyrmion with downward magnetization, with a net topological charge, Q=0, has been proven advantageous.\cite{SkrMbeg2015ground, SkrMrohart2013skyrmion, skyrMbogdanov1999stability} 

Recently, various types of devices have been proposed in which geometries play a pivotal role in control and manipulation of skyrmions to achieve desired applicability \cite{logicdevice, after27para3PhysRevApplied.13.054049, behera2020skyrmion}. A comprehensive study of magnetic skyrmion dynamics in terms of size, velocity, energy, and stability in width-varying nanotracks by micromagnetic simulations has been reported.\cite{para3aft1article} It has also been experimentally demonstrated that the transformation of stripe domain to magnetic skyrmion bubbles can be achieved using geometric constriction.\cite{film4jiang2015blowing} It has been shown that the skyrmions can be created from stripe domains using a static uniaxial strain/stress pulse at a geometrically constricted region.\cite{para3aft2} The impact of skyrmion-skyrmion and skyrmion-edge repulsion on skyrmion-based racetrack memory has been investigated.\cite{zhang2015skyrmion} Further, a pair of skyrmions can be generated from skyrmionium using spin transfer torque (STT) in a single channel device. \cite{motivationpaper} However, the deterministic generation of multi skyrmionic states has not been studied extensively. In this work, we perform a systematic study demonstrating the conversion of a skyrmionium into various skyrmion states on three different geometric designs. Also we have investigated the effect of current density to get different states of skyrmion in a single FM layer as well as in a SAF. The obtained results are interesting for interdisciplinary research on skyrmion-based neuromorphic and unconventional computing.   
\begin{figure}[h!]
\centering
\includegraphics[width=0.8\linewidth]{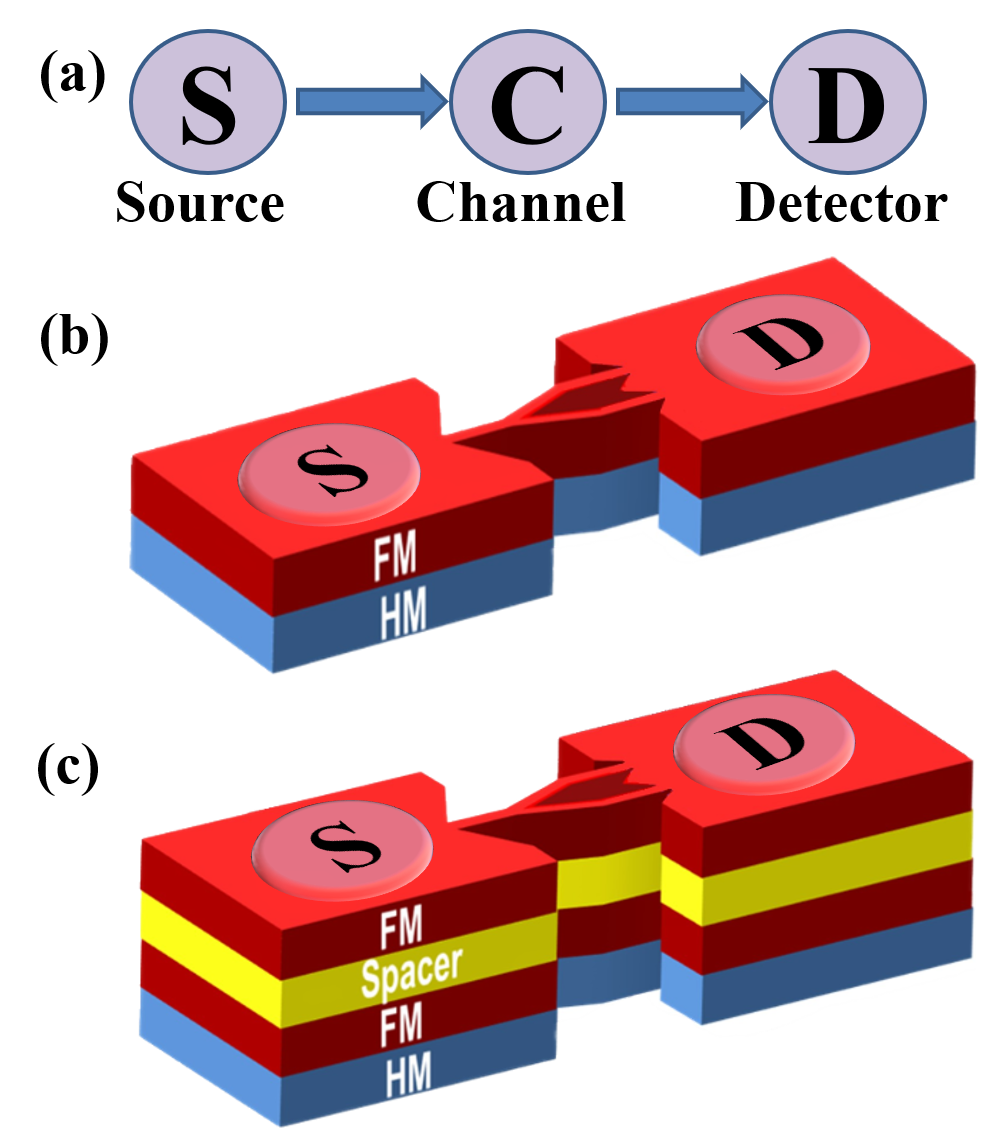}
\caption{(a)Layout of the device concept showing source (S) as input terminal, channel (C) and detector (D) as output terminal. (b) and (c) show the device schematics for a single FM layer and SAF, respectively.}
\label{fig:Fig. 1}
\end{figure}

We have performed the micromagnetic simulation using Object Oriented MicroMagnetic Framework (OOMMF)[2] with the Dzyaloshinskii-Moriya interaction (DMI) extension \cite{SkrMrohart2013skyrmion}. The schematic of the representative devices are shown in figure 1. The concept of the device, showing source (S) as input terminal, channel (C) and detector (D) as output terminal, is illustrated in (a). Fig. 1 (b) and (c) show the devices with a single ferromagnetic and a SAF structure, respectively, with specified source (input) and detector (output) terminals.  The detector is to sense the skyrmionic states and $J$ represents the current density. For the simulations, we have considered 800 nm long and 300 nm wide sample with a thickness of 0.6 nm. The width of the narrow channel is 16 nm with notches at the terminals. We considered a cell size of 2 nm x 2 nm x 0.6 nm. We have considered three types of nanostructured designs in our simulations; with 1-1 channel (Fig. 2 (a)), 1-2 channel (Fig. 2 (b)) and 1-3 channel (Fig. 2 (c)), named as Dvc-1, Dvc-2 and Dvc-3, respectively. In the simulations the current was applied in the in-plane (CIP) geometry. The modified Landau-Liftshitz-Gilbert (LLG) equation with STT term for the time-dependent spin dynamics for CIP geometry, is given by: \cite{safzhang2016magnetic, motivationpaper}
\begin{eqnarray}
\frac{d M}{d t}=-\gamma_{o} M\times H_{e f f}+\frac{\alpha}{M_{s}}\left(M\times \frac{d M}{d t}\right)\nonumber\\
+\frac{u}{M_{s}^{2}}\left(M \times \frac{\partial M}{\partial x} \times \boldsymbol{M}\right)-\frac{\beta u}{M_{s}}\left(M\times \frac{\partial M}{\partial x}\right)
\end{eqnarray}

where $M$ is the magnetization, $M_s$ is the spontaneous magnetization, $t$ is the time. The effective field is given by $H_{eff}=-\frac{1}{\mu_{O}} \frac{\partial E}{\partial M}$ and the STT coefficient is given by $u=\left|\frac{\gamma_{o} \hbar}{\mu_{o} e}\right| \frac{J P}{2 M_{S}}$. The unit polarization direction is denoted by $\hat{p}$, which is $+\hat{z}$ for nucleating skyrmionium and $+\hat{y}$ to drive the chiral structures in a forward direction. $\gamma_0$ is the gyro-magnetic ratio, $\hbar$ is reduced Planck's constant, $e$ is the electronic charge, and $\mu_0$ is the vacuum permeability constant. The first and second terms in equation (1) describe precession and damping, respectively. The third and fourth terms account for the adiabatic and non-adiabatic contribution of the spin-transfer torque (STT). The values of exchange stiffness $(A)$, Gilbert damping $(\alpha)$, spin polarization $(P)$, spontaneous magnetization $(M_s)$, DMI factor $(D)$, gyro-magnetic ratio $(\gamma)$ and perpendicular magneto-crystalline anisotropy $(K)$ are $15 pJm^{-1}$, $0.3$, $0.4$, $580 kA m^{-1}$, $3.6 mJm^{-3}$, $2.21276 \times 10^{5} m A^{-1}$ $s^{-1}$ and $0.8 MJ m^{-3}$, respectively. 

In our system, we have taken $\alpha = \beta$ for single FM layer to avoid SkHE. From an experimental point of view, an excellent way to avoid the SkHE is by using a SAF layout. We have also reproduced our results in SAF layout by considering RKKY exchange coupling for our system with an inter-layer antiferromagnetic exchange coupling value of $-2 pJ m^{-1}$. However, the coupling strength can be modulated depending upon the thickness of the spacer layer. For  the SAF system, we have considered the strength of the non-adiabatic coefficient $\beta$ of STT torque as 0.6, such that $\beta$ = 2$\alpha$,  to achieve a higher velocity at same current density.\cite{iwasaki2013current}

\begin{figure*}
\includegraphics[width=\textwidth]{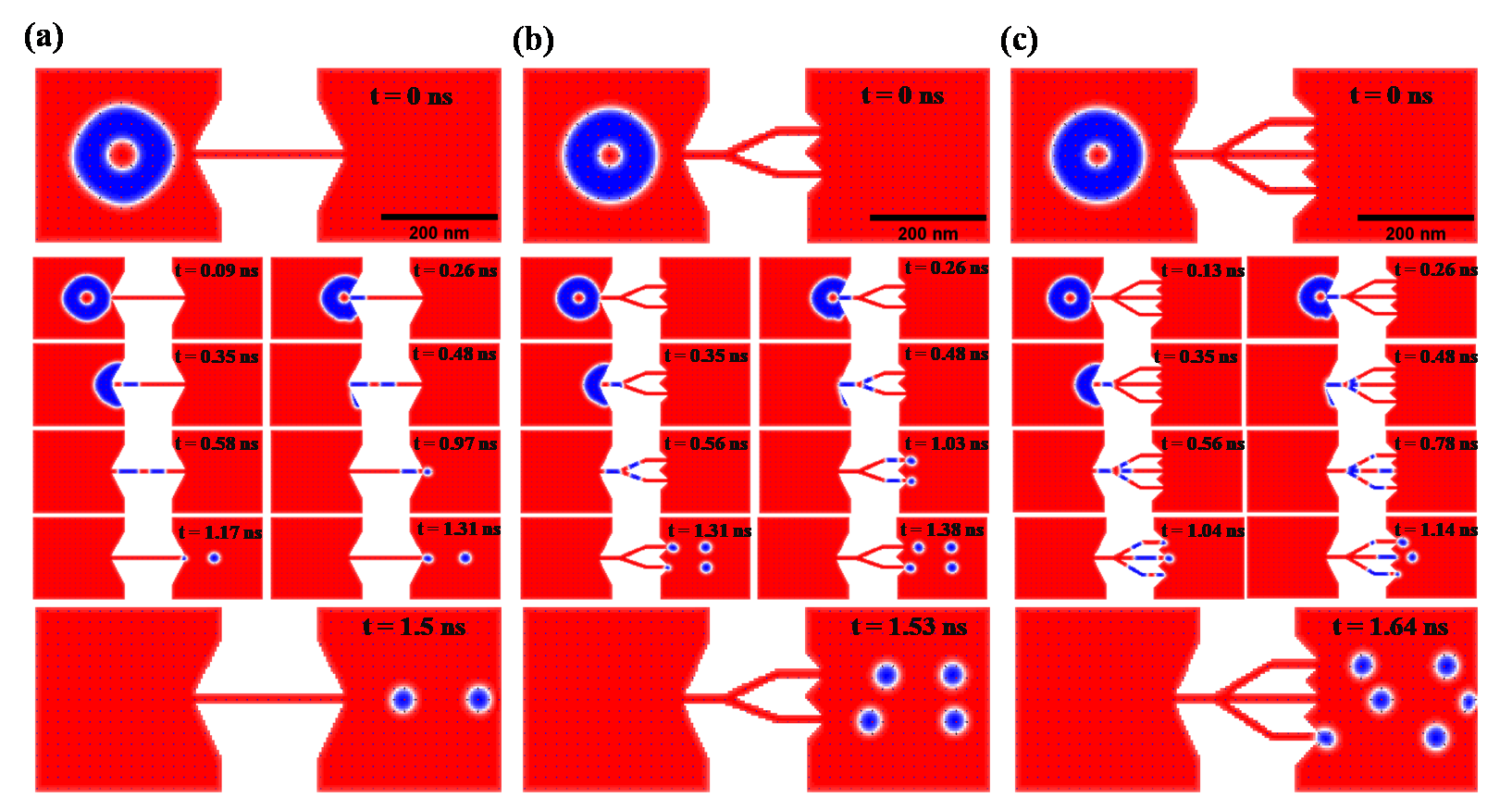}
\caption{\label{fig:wide}Snapshots of critical moments during the conversion process in (a) 1-1 channel design: Dvc-1 (b) 1-2 channel design: Dvc-2 (c) 1-3 channel design: Dvc-3 for single FM layer at $J$ = $10.5 \times 10^12$ $A m^{-2}$.}
\end{figure*}

A current density of $2.5 \times 10^6 A m^{-2}$ is applied at the S terminal of the device in an annular fashion such that the spin polarization is anti-parallel to the initial magnetic configuration. The skyrmionum, with core radius $44$ $nm$ and peripheral radius $114$ $nm$, is nucleated after the spin textures are relaxed at the input terminal of the device, as shown in fig 2 (t = 0 ns). The skyrmionium is driven through the narrow channel using STT along the current direction. As the skyrmionium reaches the narrow channel, it's peripheral spins collide at the edges (notches) and experience repulsion. Due to the applied current, the skyrmionium gets injected into the narrow channel while being converted into a domain wall. 
The spin textures of skyrmionium start to deform as it translates through the channel(s). The regions with $m_z> 0$ (red) and $m_z< 0$ (blue) is separated by the boundary with $m_z= 0$ (white). 
Two separate magnetic domains can be seen at 0.35 ns in the narrow channel(s). At 0.48 ns, the former DW pair is pushed forward and another DW pair is injected into the channel. The scope of the DW depend on the specific channel designs as they further develop according to the structure of the devices. In Dvc-1, the DW pair simply translate to the output terminal. Whereas, in case of Dvc-2 and Dvc-3, the DW pair get split at the junctions into two and three parts, respectively, which can be seen in figure 2 (b), and (c) at 0.56 ns. 

The conversion mechanism between a DW pair and a skyrmion has been already investigated in literature \cite{HYBRIDDEVICEzhou2014reversible}. In a DW pair, we can see two open parallel boundaries touching the edges of the narrow channel, where the spin directions are oriented in a way to give a topological charge Q=0.\cite{HYBRIDDEVICEzhou2014reversible} As the DW pair arrive at the end of the narrow channel they are squeezed to smaller sizes and tend to pass through the constricted terminal by the virtue of STT and repulsion\cite{thomas2012topological, p2017current} from the later batch of DW pair. When the open boundary of first DW reaches the end of the channel, it tends to move outward if the driving force (STT) is strong enough to overcome the pinning at the notch. The STT compels the second boundary to expel out to form a closed loop [as in fig. S2 ($t=1.13$ ns channel (ii)), supplementary]. The spin orientation of this loop is hedgehog and the topological charge Q=1, resulting in skyrmion. The topologically protected skyrmions float in the output terminal of the device without interruption. Alternatively, the second DW pair may get pinned at the end of the narrow channel if the driving current is not enough to expel it out [as in fig. S2 (t=3 ns), supplementary]. The critical current density for which the skyrmion ejection occurs at the end of the channel for Dvc-1, 2 and 3 are found to be $6.58 \times 10^{12}$ $A m^{-2}$, $6.85 \times 10^{12}$ $A m^{-2}$ and $7.51 \times 10^{12}$ $A m^{-2}$, respectively, in single FM. However, the current density in case of SAF is $3.75 \times 10^{12}$ $A m^{-2}$ for all three devices.

The number of skyrmions released depend on $J$. Fig 3 (a) shows the number of skyrmions as a function of current density for the HM/FM bi-layer and SAF which explains that a specific skyrmionic state is achieved corresponding to a threshold value of the current density. For FM single layer, all possible number of skyrmions have been observed for all three devices. For Dvc-1, it is observed that by tuning the $J$, 1 or 2 skyrmions are generated. Similarly, in Dvc-2 and Dvc-3, the number of skyrmions at the detector terminal are 1-4 and 1-6, respectively. In the case of SAF, due to the RKKY torque,\cite{RKKYtorquedoi} the domains in the narrow channel(s) do not get pinned at the terminal. Because of this, we observed that the ejected number of skyrmions are equal to the number of channels or twice of it. For Dvc-1\_SAF, Dvc-2\_SAF and Dvc-3\_SAF, the number of skyrmionic states are 1 or 2, 2 or  4 and 3 or 6, respectively.  It should be noted that the velocity of spin textures (skyrmionium in the source terminal, DW in narrow channel, and skyrmions in the detector terminal) in SAF is almost twice that of the corresponding single FM layer devices for a fixed value of $J$.

\begin{figure}[h!]
\centering
\includegraphics[width=1\linewidth]{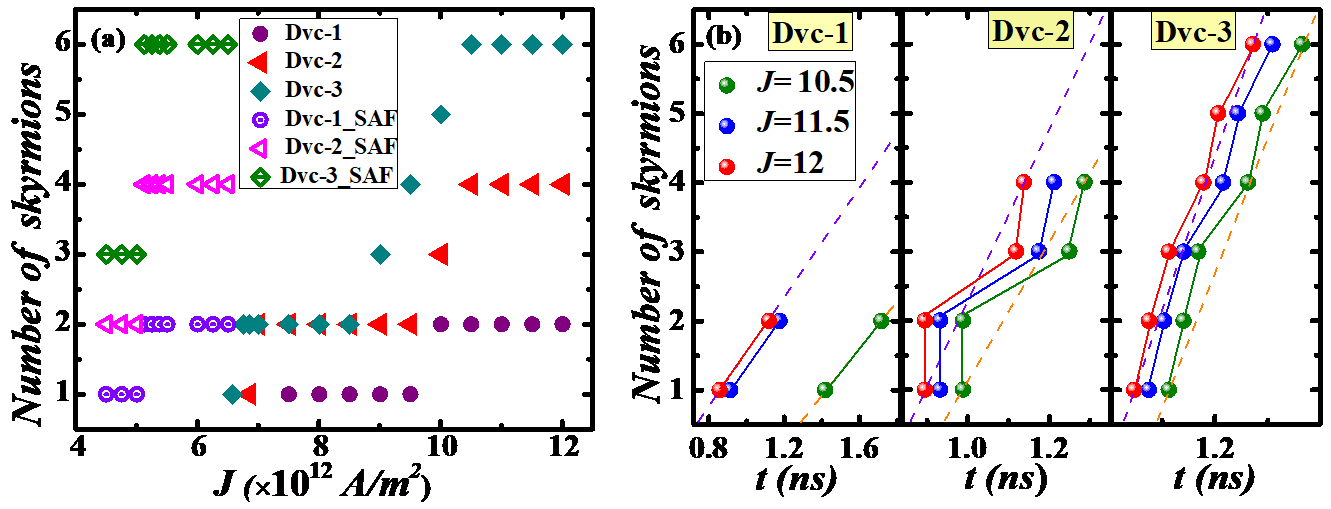}
\caption{(a) Number of skyrmions as a function of current density $J (\times 10^{12} Am^{-2}$) in single FM layer. (b) Number of skyrmions as a function of time achieved per device in single FM layer. The orange and purple dotted lines pass over the initial and final state of skyrmions released at 10.5  $\times 10^{12} Am^{-2}$ and 124  $\times 10^{12} Am^{-2}$ of current density.}
\label{fig:Fig. 3}
\end{figure}

To study the time delay between generation of each skyrmionic state, we have plotted the number of skyrmions as a function of time at $J$ $=$ $10.5 \times 10^{12}$ $A m^{-2}$ and $12 \times 10^{12}$ $A m^{-2}$, as shown in fig 3 (b) in single FM layer. We have also plotted the number of skyrmions as a function of time at different values of $J$ for single FM and SAF in fig. S3 of the supplementary information. It has been observed that faster release of the skyrmions happens at higher $J$. The delay in the release of skyrmions may be due to geometric aberrations at notch which disrupts the symmetry of the Dvc-2 and Dvc-3. Further, the pinning at the asymmetrical notch persists for a longer time. The domains running in the middle channel of Dvc-3 are released earlier than the sloped channels as they travel a shorter distance.  The time interval between release of skyrmions from the former and the later batch of domains narrows down with increasing $J$.

\begin{figure}[h!]
\centering
\includegraphics[width=1\linewidth]{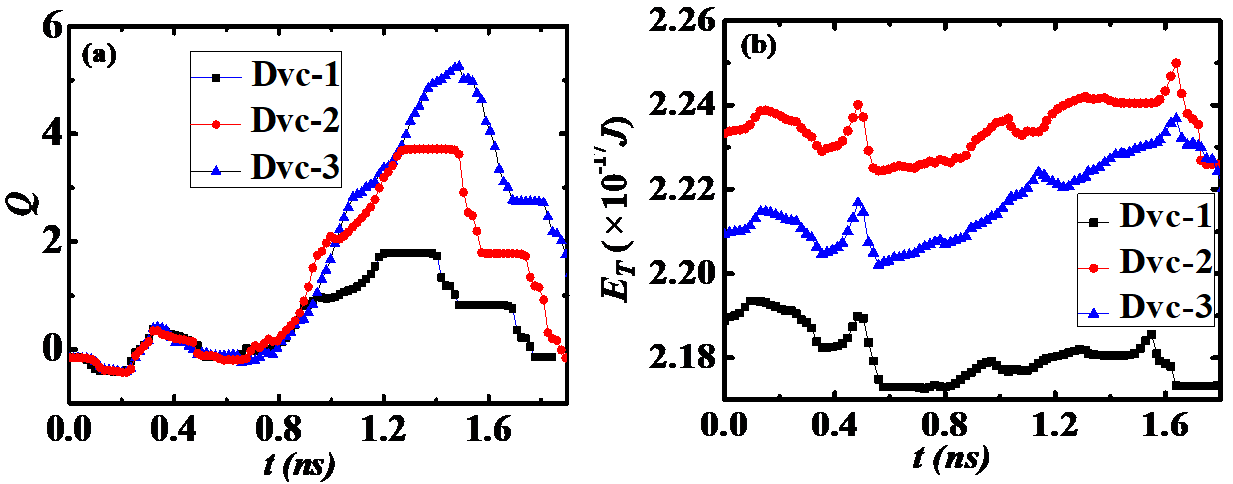}
\caption{(a) The topological number, (b) the total micromagnetic energy, (c) the demagnetization energy and (d) the DW spins as function of t is plotted for maximum states obtained in Dvc-1, Dvc-2 and Dvc-3 in single FM layer. The trend is observed to be similar in case of SAF. The DW spins are defined as the spins which have -0.2 $\leq$ m$_{z}$  $\leq$ 0.2.}
\label{fig:Fig. 4}
\end{figure}

As expected, the topological number is found to be 0 and 1 for skyrmionium and single skyrmion, respectively. The increase in topological number as a function of time corresponding to creation of skyrmionic states is shown in figure 4 (a) for maximum states (i.e., 2, 4 and 6 for Dvc-1, Dvc-2 and Dvc-3, respectively) at $J$= $10.5 \times 10^{12}$ $A m^{-2}$. The total topological numbers increase from Q=0 (for skyrmionium) to Q=+2 (for pair of skyrmions), Q=+4 (for the 4-skyrmionic states) and Q=+6 (for the 6-skyrmionic states). 
The meta-stability of the system is depicted from the fluctuations seen in the profile of total micromagnetic energy during the conversion process (see fig 4 (b)). The deformation of skyrmionium is manifested as prominent peaks in the total micromagnetic energy profile at 0.5 ns. The peaks are observed at 1.5 ns, 1.6 ns and 1.6 ns for Dvc-1, Dvc-2 and Dvc-3, respectively, when the skyrmions are formed.


To summarize, we have investigated the conversion between skyrmionium (Q=0) to skyrmion (Q=+1) in single ferromagnetic layer as well as in synthetic antiferromagnetic (SAF) system, using distinct channeled geometry to achieve multi-skyrmionic states. It is observed that by tuning the current density, the deterministic generation of skyrmions is possible. Hence, the device is more efficiently adjustable compared to other previously proposed magnetic devices, as it is easier to change the number of skyrmions. This type of designs can pave way for advancement of skyrmion-based memristors, spiking neural network (SNN), and other neuromorphic devices \cite{grollier2020neuromorphic}. Geometry-dependent generation of skyrmions enable new functionalities that may be inaccessible to conventional technologies, thus, paving way towards unconventional computing hardware resolution.


\begin{acknowledgments}
The authors thank DAE, Govt. of India and the Indo-French collaborative project supported by CEFIPRA for providing the research funding.
\end{acknowledgments}

\nocite{*}
\bibliography{reference}

\providecommand{\noopsort}[1]{}\providecommand{\singleletter}[1]{#1}%
\begin{thebibliography}{60}%
\makeatletter
\providecommand \@ifxundefined [1]{%
 \@ifx{#1\undefined}
}%
\providecommand \@ifnum [1]{%
 \ifnum #1\expandafter \@firstoftwo
 \else \expandafter \@secondoftwo
 \fi
}%
\providecommand \@ifx [1]{%
 \ifx #1\expandafter \@firstoftwo
 \else \expandafter \@secondoftwo
 \fi
}%
\providecommand \natexlab [1]{#1}%
\providecommand \enquote  [1]{``#1''}%
\providecommand \bibnamefont  [1]{#1}%
\providecommand \bibfnamefont [1]{#1}%
\providecommand \citenamefont [1]{#1}%
\providecommand \href@noop [0]{\@secondoftwo}%
\providecommand \href [0]{\begingroup \@sanitize@url \@href}%
\providecommand \@href[1]{\@@startlink{#1}\@@href}%
\providecommand \@@href[1]{\endgroup#1\@@endlink}%
\providecommand \@sanitize@url [0]{\catcode `\\12\catcode `\$12\catcode
  `\&12\catcode `\#12\catcode `\^12\catcode `\_12\catcode `\%12\relax}%
\providecommand \@@startlink[1]{}%
\providecommand \@@endlink[0]{}%
\providecommand \url  [0]{\begingroup\@sanitize@url \@url }%
\providecommand \@url [1]{\endgroup\@href {#1}{\urlprefix }}%
\providecommand \urlprefix  [0]{URL }%
\providecommand \Eprint [0]{\href }%
\providecommand \doibase [0]{https://doi.org/}%
\providecommand \selectlanguage [0]{\@gobble}%
\providecommand \bibinfo  [0]{\@secondoftwo}%
\providecommand \bibfield  [0]{\@secondoftwo}%
\providecommand \translation [1]{[#1]}%
\providecommand \BibitemOpen [0]{}%
\providecommand \bibitemStop [0]{}%
\providecommand \bibitemNoStop [0]{.\EOS\space}%
\providecommand \EOS [0]{\spacefactor3000\relax}%
\providecommand \BibitemShut  [1]{\csname bibitem#1\endcsname}%
\let\auto@bib@innerbib\@empty
\bibitem [{\citenamefont {Skyrme}(1961)}]{Skyrme4}%
  \BibitemOpen
  \bibfield  {author} {\bibinfo {author} {\bibfnamefont {T.~H.~R.}\
  \bibnamefont {Skyrme}},\ }\bibfield  {title} {\enquote {\bibinfo {title} {{A
  Nonlinear field theory}},}\ }\href {https://doi.org/10.1098/rspa.1961.0018}
  {\bibfield  {journal} {\bibinfo  {journal} {Proc. Roy. Soc. Lond. A}\
  }\textbf {\bibinfo {volume} {260}},\ \bibinfo {pages} {127--138} (\bibinfo
  {year} {1961})}\BibitemShut {NoStop}%
\bibitem [{\citenamefont {Chui}, \citenamefont {Zhou},\ and\ \citenamefont
  {Liu}(2017)}]{chui2017magnetic}%
  \BibitemOpen
  \bibfield  {author} {\bibinfo {author} {\bibfnamefont {C.-p.}\ \bibnamefont
  {Chui}}, \bibinfo {author} {\bibfnamefont {Y.}~\bibnamefont {Zhou}},\ and\
  \bibinfo {author} {\bibfnamefont {W.}~\bibnamefont {Liu}},\ }\bibfield
  {title} {\enquote {\bibinfo {title} {Magnetic skyrmions as information
  carriers},}\ }\href@noop {} {\bibfield  {journal} {\bibinfo  {journal}
  {TECHNOLOGY}\ }\textbf {\bibinfo {volume} {19}},\ \bibinfo {pages} {18}
  (\bibinfo {year} {2017})}\BibitemShut {NoStop}%
\bibitem [{\citenamefont {Wang}\ \emph {et~al.}(2015)\citenamefont {Wang},
  \citenamefont {Beg}, \citenamefont {Zhang}, \citenamefont {Kuch},\ and\
  \citenamefont {Fangohr}}]{microwavedevice}%
  \BibitemOpen
  \bibfield  {author} {\bibinfo {author} {\bibfnamefont {W.}~\bibnamefont
  {Wang}}, \bibinfo {author} {\bibfnamefont {M.}~\bibnamefont {Beg}}, \bibinfo
  {author} {\bibfnamefont {B.}~\bibnamefont {Zhang}}, \bibinfo {author}
  {\bibfnamefont {W.}~\bibnamefont {Kuch}},\ and\ \bibinfo {author}
  {\bibfnamefont {H.}~\bibnamefont {Fangohr}},\ }\bibfield  {title} {\enquote
  {\bibinfo {title} {Driving magnetic skyrmions with microwave fields},}\
  }\href@noop {} {\bibfield  {journal} {\bibinfo  {journal} {Physical Review
  B}\ }\textbf {\bibinfo {volume} {92}},\ \bibinfo {pages} {020403} (\bibinfo
  {year} {2015})}\BibitemShut {NoStop}%
\bibitem [{\citenamefont {Xia}\ \emph {et~al.}(2017)\citenamefont {Xia},
  \citenamefont {Huang}, \citenamefont {Zhang}, \citenamefont {Kang},
  \citenamefont {Zheng}, \citenamefont {Liu}, \citenamefont {Zhao},\ and\
  \citenamefont {Zhou}}]{microwave1}%
  \BibitemOpen
  \bibfield  {author} {\bibinfo {author} {\bibfnamefont {J.}~\bibnamefont
  {Xia}}, \bibinfo {author} {\bibfnamefont {Y.}~\bibnamefont {Huang}}, \bibinfo
  {author} {\bibfnamefont {X.}~\bibnamefont {Zhang}}, \bibinfo {author}
  {\bibfnamefont {W.}~\bibnamefont {Kang}}, \bibinfo {author} {\bibfnamefont
  {C.}~\bibnamefont {Zheng}}, \bibinfo {author} {\bibfnamefont
  {X.}~\bibnamefont {Liu}}, \bibinfo {author} {\bibfnamefont {W.}~\bibnamefont
  {Zhao}},\ and\ \bibinfo {author} {\bibfnamefont {Y.}~\bibnamefont {Zhou}},\
  }\bibfield  {title} {\enquote {\bibinfo {title} {A microwave field-driven
  transistor-like skyrmionic device with the microwave current-assisted
  skyrmion creation},}\ }\href {https://doi.org/10.1063/1.4999013} {\bibfield
  {journal} {\bibinfo  {journal} {Journal of Applied Physics}\ }\textbf
  {\bibinfo {volume} {122}},\ \bibinfo {pages} {153901} (\bibinfo {year}
  {2017})}\BibitemShut {NoStop}%
\bibitem [{\citenamefont {Zhang}\ \emph
  {et~al.}(2015{\natexlab{a}})\citenamefont {Zhang}, \citenamefont {Ezawa},
  \citenamefont {Xiao}, \citenamefont {Zhao}, \citenamefont {Liu},\ and\
  \citenamefont {Zhou}}]{magnondevice}%
  \BibitemOpen
  \bibfield  {author} {\bibinfo {author} {\bibfnamefont {X.}~\bibnamefont
  {Zhang}}, \bibinfo {author} {\bibfnamefont {M.}~\bibnamefont {Ezawa}},
  \bibinfo {author} {\bibfnamefont {D.}~\bibnamefont {Xiao}}, \bibinfo {author}
  {\bibfnamefont {G.}~\bibnamefont {Zhao}}, \bibinfo {author} {\bibfnamefont
  {Y.}~\bibnamefont {Liu}},\ and\ \bibinfo {author} {\bibfnamefont
  {Y.}~\bibnamefont {Zhou}},\ }\bibfield  {title} {\enquote {\bibinfo {title}
  {All-magnetic control of skyrmions in nanowires by a spin wave},}\
  }\href@noop {} {\bibfield  {journal} {\bibinfo  {journal} {Nanotechnology}\
  }\textbf {\bibinfo {volume} {26}},\ \bibinfo {pages} {225701} (\bibinfo
  {year} {2015}{\natexlab{a}})}\BibitemShut {NoStop}%
\bibitem [{\citenamefont {Xing}, \citenamefont {Zhou},\ and\ \citenamefont
  {Braun}(2020)}]{magnondevice1}%
  \BibitemOpen
  \bibfield  {author} {\bibinfo {author} {\bibfnamefont {X.}~\bibnamefont
  {Xing}}, \bibinfo {author} {\bibfnamefont {Y.}~\bibnamefont {Zhou}},\ and\
  \bibinfo {author} {\bibfnamefont {H.}~\bibnamefont {Braun}},\ }\bibfield
  {title} {\enquote {\bibinfo {title} {Magnetic skyrmion tubes as nonplanar
  magnonic waveguides},}\ }\href
  {https://doi.org/10.1103/PhysRevApplied.13.034051} {\bibfield  {journal}
  {\bibinfo  {journal} {Phys. Rev. Applied}\ }\textbf {\bibinfo {volume}
  {13}},\ \bibinfo {pages} {034051} (\bibinfo {year} {2020})}\BibitemShut
  {NoStop}%
\bibitem [{\citenamefont {Chauwin}\ \emph {et~al.}(2019)\citenamefont
  {Chauwin}, \citenamefont {Hu}, \citenamefont {Garcia-Sanchez}, \citenamefont
  {Betrabet}, \citenamefont {Paler}, \citenamefont {Moutafis},\ and\
  \citenamefont {Friedman}}]{Quantumcomputing}%
  \BibitemOpen
  \bibfield  {author} {\bibinfo {author} {\bibfnamefont {M.}~\bibnamefont
  {Chauwin}}, \bibinfo {author} {\bibfnamefont {X.}~\bibnamefont {Hu}},
  \bibinfo {author} {\bibfnamefont {F.}~\bibnamefont {Garcia-Sanchez}},
  \bibinfo {author} {\bibfnamefont {N.}~\bibnamefont {Betrabet}}, \bibinfo
  {author} {\bibfnamefont {A.}~\bibnamefont {Paler}}, \bibinfo {author}
  {\bibfnamefont {C.}~\bibnamefont {Moutafis}},\ and\ \bibinfo {author}
  {\bibfnamefont {J.~S.}\ \bibnamefont {Friedman}},\ }\bibfield  {title}
  {\enquote {\bibinfo {title} {Skyrmion logic system for large-scale reversible
  computation},}\ }\href {https://doi.org/10.1103/PhysRevApplied.12.064053}
  {\bibfield  {journal} {\bibinfo  {journal} {Phys. Rev. Applied}\ }\textbf
  {\bibinfo {volume} {12}},\ \bibinfo {pages} {064053} (\bibinfo {year}
  {2019})}\BibitemShut {NoStop}%
\bibitem [{\citenamefont {Li}\ \emph {et~al.}(2021)\citenamefont {Li},
  \citenamefont {Kang}, \citenamefont {Zhang}, \citenamefont {Nie},
  \citenamefont {Zhou}, \citenamefont {Wang},\ and\ \citenamefont
  {Zhao}}]{li2021magnetic}%
  \BibitemOpen
  \bibfield  {author} {\bibinfo {author} {\bibfnamefont {S.}~\bibnamefont
  {Li}}, \bibinfo {author} {\bibfnamefont {W.}~\bibnamefont {Kang}}, \bibinfo
  {author} {\bibfnamefont {X.}~\bibnamefont {Zhang}}, \bibinfo {author}
  {\bibfnamefont {T.}~\bibnamefont {Nie}}, \bibinfo {author} {\bibfnamefont
  {Y.}~\bibnamefont {Zhou}}, \bibinfo {author} {\bibfnamefont {K.~L.}\
  \bibnamefont {Wang}},\ and\ \bibinfo {author} {\bibfnamefont
  {W.}~\bibnamefont {Zhao}},\ }\bibfield  {title} {\enquote {\bibinfo {title}
  {Magnetic skyrmions for unconventional computing},}\ }\href@noop {}
  {\bibfield  {journal} {\bibinfo  {journal} {Materials Horizons}\ }\textbf
  {\bibinfo {volume} {8}},\ \bibinfo {pages} {854--868} (\bibinfo {year}
  {2021})}\BibitemShut {NoStop}%
\bibitem [{\citenamefont {Zhang}, \citenamefont {Ezawa},\ and\ \citenamefont
  {Zhou}(2015)}]{logicdevice}%
  \BibitemOpen
  \bibfield  {author} {\bibinfo {author} {\bibfnamefont {X.}~\bibnamefont
  {Zhang}}, \bibinfo {author} {\bibfnamefont {M.}~\bibnamefont {Ezawa}},\ and\
  \bibinfo {author} {\bibfnamefont {Y.}~\bibnamefont {Zhou}},\ }\bibfield
  {title} {\enquote {\bibinfo {title} {Magnetic skyrmion logic gates:
  conversion, duplication and merging of skyrmions},}\ }\href@noop {}
  {\bibfield  {journal} {\bibinfo  {journal} {Scientific Reports}\ }\textbf
  {\bibinfo {volume} {5}},\ \bibinfo {pages} {1--8} (\bibinfo {year}
  {2015})}\BibitemShut {NoStop}%
\bibitem [{\citenamefont {Luo}\ \emph {et~al.}(2018)\citenamefont {Luo},
  \citenamefont {Song}, \citenamefont {Li}, \citenamefont {Zhang},
  \citenamefont {Hong}, \citenamefont {Yang}, \citenamefont {Zou},
  \citenamefont {Xu},\ and\ \citenamefont {You}}]{logicdevice1}%
  \BibitemOpen
  \bibfield  {author} {\bibinfo {author} {\bibfnamefont {S.}~\bibnamefont
  {Luo}}, \bibinfo {author} {\bibfnamefont {M.}~\bibnamefont {Song}}, \bibinfo
  {author} {\bibfnamefont {X.}~\bibnamefont {Li}}, \bibinfo {author}
  {\bibfnamefont {Y.}~\bibnamefont {Zhang}}, \bibinfo {author} {\bibfnamefont
  {J.}~\bibnamefont {Hong}}, \bibinfo {author} {\bibfnamefont {X.}~\bibnamefont
  {Yang}}, \bibinfo {author} {\bibfnamefont {X.}~\bibnamefont {Zou}}, \bibinfo
  {author} {\bibfnamefont {N.}~\bibnamefont {Xu}},\ and\ \bibinfo {author}
  {\bibfnamefont {L.}~\bibnamefont {You}},\ }\bibfield  {title} {\enquote
  {\bibinfo {title} {Reconfigurable skyrmion logic gates},}\ }\href@noop {}
  {\bibfield  {journal} {\bibinfo  {journal} {Nano letters}\ }\textbf {\bibinfo
  {volume} {18}},\ \bibinfo {pages} {1180--1184} (\bibinfo {year}
  {2018})}\BibitemShut {NoStop}%
\bibitem [{\citenamefont {Kiselev}\ \emph {et~al.}(2011)\citenamefont
  {Kiselev}, \citenamefont {Bogdanov}, \citenamefont {Sch{\"a}fer},\ and\
  \citenamefont {R{\"o}{\ss}ler}}]{storage}%
  \BibitemOpen
  \bibfield  {author} {\bibinfo {author} {\bibfnamefont {N.}~\bibnamefont
  {Kiselev}}, \bibinfo {author} {\bibfnamefont {A.}~\bibnamefont {Bogdanov}},
  \bibinfo {author} {\bibfnamefont {R.}~\bibnamefont {Sch{\"a}fer}},\ and\
  \bibinfo {author} {\bibfnamefont {U.}~\bibnamefont {R{\"o}{\ss}ler}},\
  }\bibfield  {title} {\enquote {\bibinfo {title} {Chiral skyrmions in thin
  magnetic films: new objects for magnetic storage technologies?}}\ }\href@noop
  {} {\bibfield  {journal} {\bibinfo  {journal} {Journal of Physics D: Applied
  Physics}\ }\textbf {\bibinfo {volume} {44}},\ \bibinfo {pages} {392001}
  (\bibinfo {year} {2011})}\BibitemShut {NoStop}%
\bibitem [{\citenamefont {Bessarab}\ \emph {et~al.}(2018)\citenamefont
  {Bessarab}, \citenamefont {M{\"u}ller}, \citenamefont {Lobanov},
  \citenamefont {Rybakov}, \citenamefont {Kiselev}, \citenamefont
  {J{\'o}nsson}, \citenamefont {Uzdin}, \citenamefont {Bl{\"u}gel},
  \citenamefont {Bergqvist},\ and\ \citenamefont {Delin}}]{storage1}%
  \BibitemOpen
  \bibfield  {author} {\bibinfo {author} {\bibfnamefont {P.~F.}\ \bibnamefont
  {Bessarab}}, \bibinfo {author} {\bibfnamefont {G.~P.}\ \bibnamefont
  {M{\"u}ller}}, \bibinfo {author} {\bibfnamefont {I.~S.}\ \bibnamefont
  {Lobanov}}, \bibinfo {author} {\bibfnamefont {F.~N.}\ \bibnamefont
  {Rybakov}}, \bibinfo {author} {\bibfnamefont {N.~S.}\ \bibnamefont
  {Kiselev}}, \bibinfo {author} {\bibfnamefont {H.}~\bibnamefont
  {J{\'o}nsson}}, \bibinfo {author} {\bibfnamefont {V.~M.}\ \bibnamefont
  {Uzdin}}, \bibinfo {author} {\bibfnamefont {S.}~\bibnamefont {Bl{\"u}gel}},
  \bibinfo {author} {\bibfnamefont {L.}~\bibnamefont {Bergqvist}},\ and\
  \bibinfo {author} {\bibfnamefont {A.}~\bibnamefont {Delin}},\ }\bibfield
  {title} {\enquote {\bibinfo {title} {Lifetime of racetrack skyrmions},}\
  }\href@noop {} {\bibfield  {journal} {\bibinfo  {journal} {Scientific
  Reports}\ }\textbf {\bibinfo {volume} {8}},\ \bibinfo {pages} {1--10}
  (\bibinfo {year} {2018})}\BibitemShut {NoStop}%
\bibitem [{\citenamefont {Everschor-Sitte}, \citenamefont {Sinova},\ and\
  \citenamefont {Abanov}(2018)}]{storage2}%
  \BibitemOpen
  \bibfield  {author} {\bibinfo {author} {\bibfnamefont {K.}~\bibnamefont
  {Everschor-Sitte}}, \bibinfo {author} {\bibfnamefont {J.}~\bibnamefont
  {Sinova}},\ and\ \bibinfo {author} {\bibfnamefont {A.}~\bibnamefont
  {Abanov}},\ }\bibfield  {title} {\enquote {\bibinfo {title} {Painting and
  erasing skyrmions},}\ }\href@noop {} {\bibfield  {journal} {\bibinfo
  {journal} {Nature Electronics}\ }\textbf {\bibinfo {volume} {1}},\ \bibinfo
  {pages} {266--267} (\bibinfo {year} {2018})}\BibitemShut {NoStop}%
\bibitem [{\citenamefont {Fert}, \citenamefont {Reyren},\ and\ \citenamefont
  {Cros}(2017)}]{storage3}%
  \BibitemOpen
  \bibfield  {author} {\bibinfo {author} {\bibfnamefont {A.}~\bibnamefont
  {Fert}}, \bibinfo {author} {\bibfnamefont {N.}~\bibnamefont {Reyren}},\ and\
  \bibinfo {author} {\bibfnamefont {V.}~\bibnamefont {Cros}},\ }\bibfield
  {title} {\enquote {\bibinfo {title} {Magnetic skyrmions: advances in physics
  and potential applications},}\ }\href@noop {} {\bibfield  {journal} {\bibinfo
   {journal} {Nature Reviews Materials}\ }\textbf {\bibinfo {volume} {2}},\
  \bibinfo {pages} {1--15} (\bibinfo {year} {2017})}\BibitemShut {NoStop}%
\bibitem [{\citenamefont {M{\"u}ller}(2017)}]{storage4}%
  \BibitemOpen
  \bibfield  {author} {\bibinfo {author} {\bibfnamefont {J.}~\bibnamefont
  {M{\"u}ller}},\ }\bibfield  {title} {\enquote {\bibinfo {title} {Magnetic
  skyrmions on a two-lane racetrack},}\ }\href@noop {} {\bibfield  {journal}
  {\bibinfo  {journal} {New Journal of Physics}\ }\textbf {\bibinfo {volume}
  {19}},\ \bibinfo {pages} {025002} (\bibinfo {year} {2017})}\BibitemShut
  {NoStop}%
\bibitem [{\citenamefont {Thiaville}\ \emph {et~al.}(2012)\citenamefont
  {Thiaville}, \citenamefont {Rohart}, \citenamefont {Ju{\'e}}, \citenamefont
  {Cros},\ and\ \citenamefont {Fert}}]{film1}%
  \BibitemOpen
  \bibfield  {author} {\bibinfo {author} {\bibfnamefont {A.}~\bibnamefont
  {Thiaville}}, \bibinfo {author} {\bibfnamefont {S.}~\bibnamefont {Rohart}},
  \bibinfo {author} {\bibfnamefont {{\'E}.}~\bibnamefont {Ju{\'e}}}, \bibinfo
  {author} {\bibfnamefont {V.}~\bibnamefont {Cros}},\ and\ \bibinfo {author}
  {\bibfnamefont {A.}~\bibnamefont {Fert}},\ }\bibfield  {title} {\enquote
  {\bibinfo {title} {Dynamics of dzyaloshinskii domain walls in ultrathin
  magnetic films},}\ }\href@noop {} {\bibfield  {journal} {\bibinfo  {journal}
  {EPL (Europhysics Letters)}\ }\textbf {\bibinfo {volume} {100}},\ \bibinfo
  {pages} {57002} (\bibinfo {year} {2012})}\BibitemShut {NoStop}%
\bibitem [{\citenamefont {Jiang}\ \emph
  {et~al.}(2017{\natexlab{a}})\citenamefont {Jiang}, \citenamefont {Chen},
  \citenamefont {Liu}, \citenamefont {Zang}, \citenamefont {Te~Velthuis},\ and\
  \citenamefont {Hoffmann}}]{jiang2017skyrmions}%
  \BibitemOpen
  \bibfield  {author} {\bibinfo {author} {\bibfnamefont {W.}~\bibnamefont
  {Jiang}}, \bibinfo {author} {\bibfnamefont {G.}~\bibnamefont {Chen}},
  \bibinfo {author} {\bibfnamefont {K.}~\bibnamefont {Liu}}, \bibinfo {author}
  {\bibfnamefont {J.}~\bibnamefont {Zang}}, \bibinfo {author} {\bibfnamefont
  {S.~G.}\ \bibnamefont {Te~Velthuis}},\ and\ \bibinfo {author} {\bibfnamefont
  {A.}~\bibnamefont {Hoffmann}},\ }\bibfield  {title} {\enquote {\bibinfo
  {title} {Skyrmions in magnetic multilayers},}\ }\href@noop {} {\bibfield
  {journal} {\bibinfo  {journal} {Physics Reports}\ }\textbf {\bibinfo {volume}
  {704}},\ \bibinfo {pages} {1--49} (\bibinfo {year}
  {2017}{\natexlab{a}})}\BibitemShut {NoStop}%
\bibitem [{\citenamefont {Woo}\ \emph {et~al.}(2016)\citenamefont {Woo},
  \citenamefont {Litzius}, \citenamefont {Kr{\"u}ger}, \citenamefont {Im},
  \citenamefont {Caretta}, \citenamefont {Richter}, \citenamefont {Mann},
  \citenamefont {Krone}, \citenamefont {Reeve}, \citenamefont {Weigand} \emph
  {et~al.}}]{film3}%
  \BibitemOpen
  \bibfield  {author} {\bibinfo {author} {\bibfnamefont {S.}~\bibnamefont
  {Woo}}, \bibinfo {author} {\bibfnamefont {K.}~\bibnamefont {Litzius}},
  \bibinfo {author} {\bibfnamefont {B.}~\bibnamefont {Kr{\"u}ger}}, \bibinfo
  {author} {\bibfnamefont {M.-Y.}\ \bibnamefont {Im}}, \bibinfo {author}
  {\bibfnamefont {L.}~\bibnamefont {Caretta}}, \bibinfo {author} {\bibfnamefont
  {K.}~\bibnamefont {Richter}}, \bibinfo {author} {\bibfnamefont
  {M.}~\bibnamefont {Mann}}, \bibinfo {author} {\bibfnamefont {A.}~\bibnamefont
  {Krone}}, \bibinfo {author} {\bibfnamefont {R.~M.}\ \bibnamefont {Reeve}},
  \bibinfo {author} {\bibfnamefont {M.}~\bibnamefont {Weigand}}, \emph
  {et~al.},\ }\bibfield  {title} {\enquote {\bibinfo {title} {Observation of
  room-temperature magnetic skyrmions and their current-driven dynamics in
  ultrathin metallic ferromagnets},}\ }\href@noop {} {\bibfield  {journal}
  {\bibinfo  {journal} {Nature Materials}\ }\textbf {\bibinfo {volume} {15}},\
  \bibinfo {pages} {501--506} (\bibinfo {year} {2016})}\BibitemShut {NoStop}%
\bibitem [{\citenamefont {Jiang}\ \emph {et~al.}(2015)\citenamefont {Jiang},
  \citenamefont {Upadhyaya}, \citenamefont {Zhang}, \citenamefont {Yu},
  \citenamefont {Jungfleisch}, \citenamefont {Fradin}, \citenamefont {Pearson},
  \citenamefont {Tserkovnyak}, \citenamefont {Wang}, \citenamefont {Heinonen}
  \emph {et~al.}}]{film4jiang2015blowing}%
  \BibitemOpen
  \bibfield  {author} {\bibinfo {author} {\bibfnamefont {W.}~\bibnamefont
  {Jiang}}, \bibinfo {author} {\bibfnamefont {P.}~\bibnamefont {Upadhyaya}},
  \bibinfo {author} {\bibfnamefont {W.}~\bibnamefont {Zhang}}, \bibinfo
  {author} {\bibfnamefont {G.}~\bibnamefont {Yu}}, \bibinfo {author}
  {\bibfnamefont {M.~B.}\ \bibnamefont {Jungfleisch}}, \bibinfo {author}
  {\bibfnamefont {F.~Y.}\ \bibnamefont {Fradin}}, \bibinfo {author}
  {\bibfnamefont {J.~E.}\ \bibnamefont {Pearson}}, \bibinfo {author}
  {\bibfnamefont {Y.}~\bibnamefont {Tserkovnyak}}, \bibinfo {author}
  {\bibfnamefont {K.~L.}\ \bibnamefont {Wang}}, \bibinfo {author}
  {\bibfnamefont {O.}~\bibnamefont {Heinonen}}, \emph {et~al.},\ }\bibfield
  {title} {\enquote {\bibinfo {title} {Blowing magnetic skyrmion bubbles},}\
  }\href@noop {} {\bibfield  {journal} {\bibinfo  {journal} {Science}\ }\textbf
  {\bibinfo {volume} {349}},\ \bibinfo {pages} {283--286} (\bibinfo {year}
  {2015})}\BibitemShut {NoStop}%
\bibitem [{\citenamefont {Everschor-Sitte}\ \emph {et~al.}(2018)\citenamefont
  {Everschor-Sitte}, \citenamefont {Masell}, \citenamefont {Reeve},\ and\
  \citenamefont {Kl{\"a}ui}}]{film5everschor2018perspective}%
  \BibitemOpen
  \bibfield  {author} {\bibinfo {author} {\bibfnamefont {K.}~\bibnamefont
  {Everschor-Sitte}}, \bibinfo {author} {\bibfnamefont {J.}~\bibnamefont
  {Masell}}, \bibinfo {author} {\bibfnamefont {R.~M.}\ \bibnamefont {Reeve}},\
  and\ \bibinfo {author} {\bibfnamefont {M.}~\bibnamefont {Kl{\"a}ui}},\
  }\bibfield  {title} {\enquote {\bibinfo {title} {Perspective: Magnetic
  skyrmions—overview of recent progress in an active research field},}\
  }\href@noop {} {\bibfield  {journal} {\bibinfo  {journal} {Journal of Applied
  Physics}\ }\textbf {\bibinfo {volume} {124}},\ \bibinfo {pages} {240901}
  (\bibinfo {year} {2018})}\BibitemShut {NoStop}%
\bibitem [{\citenamefont {Heinze}\ \emph {et~al.}(2011)\citenamefont {Heinze},
  \citenamefont {Von~Bergmann}, \citenamefont {Menzel}, \citenamefont {Brede},
  \citenamefont {Kubetzka}, \citenamefont {Wiesendanger}, \citenamefont
  {Bihlmayer},\ and\ \citenamefont {Bl{\"u}gel}}]{heinze2011spontaneous}%
  \BibitemOpen
  \bibfield  {author} {\bibinfo {author} {\bibfnamefont {S.}~\bibnamefont
  {Heinze}}, \bibinfo {author} {\bibfnamefont {K.}~\bibnamefont
  {Von~Bergmann}}, \bibinfo {author} {\bibfnamefont {M.}~\bibnamefont
  {Menzel}}, \bibinfo {author} {\bibfnamefont {J.}~\bibnamefont {Brede}},
  \bibinfo {author} {\bibfnamefont {A.}~\bibnamefont {Kubetzka}}, \bibinfo
  {author} {\bibfnamefont {R.}~\bibnamefont {Wiesendanger}}, \bibinfo {author}
  {\bibfnamefont {G.}~\bibnamefont {Bihlmayer}},\ and\ \bibinfo {author}
  {\bibfnamefont {S.}~\bibnamefont {Bl{\"u}gel}},\ }\bibfield  {title}
  {\enquote {\bibinfo {title} {Spontaneous atomic-scale magnetic skyrmion
  lattice in two dimensions},}\ }\href@noop {} {\bibfield  {journal} {\bibinfo
  {journal} {Nature Physics}\ }\textbf {\bibinfo {volume} {7}},\ \bibinfo
  {pages} {713--718} (\bibinfo {year} {2011})}\BibitemShut {NoStop}%
\bibitem [{\citenamefont {Okubo}, \citenamefont {Chung},\ and\ \citenamefont
  {Kawamura}(2012)}]{okubo2012multiple}%
  \BibitemOpen
  \bibfield  {author} {\bibinfo {author} {\bibfnamefont {T.}~\bibnamefont
  {Okubo}}, \bibinfo {author} {\bibfnamefont {S.}~\bibnamefont {Chung}},\ and\
  \bibinfo {author} {\bibfnamefont {H.}~\bibnamefont {Kawamura}},\ }\bibfield
  {title} {\enquote {\bibinfo {title} {Multiple-q states and the skyrmion
  lattice of the triangular-lattice heisenberg antiferromagnet under magnetic
  fields},}\ }\href@noop {} {\bibfield  {journal} {\bibinfo  {journal}
  {Physical Review Letters}\ }\textbf {\bibinfo {volume} {108}},\ \bibinfo
  {pages} {017206} (\bibinfo {year} {2012})}\BibitemShut {NoStop}%
\bibitem [{\citenamefont {Montoya}\ \emph {et~al.}(2017)\citenamefont
  {Montoya}, \citenamefont {Couture}, \citenamefont {Chess}, \citenamefont
  {Lee}, \citenamefont {Kent}, \citenamefont {Henze}, \citenamefont {Sinha},
  \citenamefont {Im}, \citenamefont {Kevan}, \citenamefont {Fischer} \emph
  {et~al.}}]{montoya2017tailoring}%
  \BibitemOpen
  \bibfield  {author} {\bibinfo {author} {\bibfnamefont {S.}~\bibnamefont
  {Montoya}}, \bibinfo {author} {\bibfnamefont {S.}~\bibnamefont {Couture}},
  \bibinfo {author} {\bibfnamefont {J.}~\bibnamefont {Chess}}, \bibinfo
  {author} {\bibfnamefont {J.}~\bibnamefont {Lee}}, \bibinfo {author}
  {\bibfnamefont {N.}~\bibnamefont {Kent}}, \bibinfo {author} {\bibfnamefont
  {D.}~\bibnamefont {Henze}}, \bibinfo {author} {\bibfnamefont
  {S.}~\bibnamefont {Sinha}}, \bibinfo {author} {\bibfnamefont {M.-Y.}\
  \bibnamefont {Im}}, \bibinfo {author} {\bibfnamefont {S.}~\bibnamefont
  {Kevan}}, \bibinfo {author} {\bibfnamefont {P.}~\bibnamefont {Fischer}},
  \emph {et~al.},\ }\bibfield  {title} {\enquote {\bibinfo {title} {Tailoring
  magnetic energies to form dipole skyrmions and skyrmion lattices},}\
  }\href@noop {} {\bibfield  {journal} {\bibinfo  {journal} {Physical Review
  B}\ }\textbf {\bibinfo {volume} {95}},\ \bibinfo {pages} {024415} (\bibinfo
  {year} {2017})}\BibitemShut {NoStop}%
\bibitem [{\citenamefont {Tokura}\ and\ \citenamefont
  {Kanazawa}(2020)}]{tokura2020magnetic}%
  \BibitemOpen
  \bibfield  {author} {\bibinfo {author} {\bibfnamefont {Y.}~\bibnamefont
  {Tokura}}\ and\ \bibinfo {author} {\bibfnamefont {N.}~\bibnamefont
  {Kanazawa}},\ }\bibfield  {title} {\enquote {\bibinfo {title} {Magnetic
  skyrmion materials},}\ }\href@noop {} {\bibfield  {journal} {\bibinfo
  {journal} {Chemical Reviews}\ }\textbf {\bibinfo {volume} {121}},\ \bibinfo
  {pages} {2857--2897} (\bibinfo {year} {2020})}\BibitemShut {NoStop}%
\bibitem [{\citenamefont {Garel}\ and\ \citenamefont
  {Doniach}(1982)}]{garel1982phase}%
  \BibitemOpen
  \bibfield  {author} {\bibinfo {author} {\bibfnamefont {T.}~\bibnamefont
  {Garel}}\ and\ \bibinfo {author} {\bibfnamefont {S.}~\bibnamefont
  {Doniach}},\ }\bibfield  {title} {\enquote {\bibinfo {title} {Phase
  transitions with spontaneous modulation-the dipolar ising ferromagnet},}\
  }\href@noop {} {\bibfield  {journal} {\bibinfo  {journal} {Physical Review
  B}\ }\textbf {\bibinfo {volume} {26}},\ \bibinfo {pages} {325} (\bibinfo
  {year} {1982})}\BibitemShut {NoStop}%
\bibitem [{\citenamefont {Lin}, \citenamefont {Grundy},\ and\ \citenamefont
  {Giess}(1973)}]{lin1973bubble}%
  \BibitemOpen
  \bibfield  {author} {\bibinfo {author} {\bibfnamefont {Y.}~\bibnamefont
  {Lin}}, \bibinfo {author} {\bibfnamefont {P.}~\bibnamefont {Grundy}},\ and\
  \bibinfo {author} {\bibfnamefont {E.}~\bibnamefont {Giess}},\ }\bibfield
  {title} {\enquote {\bibinfo {title} {Bubble domains in magnetostatically
  coupled garnet films},}\ }\href@noop {} {\bibfield  {journal} {\bibinfo
  {journal} {Applied Physics Letters}\ }\textbf {\bibinfo {volume} {23}},\
  \bibinfo {pages} {485--487} (\bibinfo {year} {1973})}\BibitemShut {NoStop}%
\bibitem [{\citenamefont {Nagaosa}\ and\ \citenamefont
  {Tokura}(2013)}]{nagaosa2013topological}%
  \BibitemOpen
  \bibfield  {author} {\bibinfo {author} {\bibfnamefont {N.}~\bibnamefont
  {Nagaosa}}\ and\ \bibinfo {author} {\bibfnamefont {Y.}~\bibnamefont
  {Tokura}},\ }\bibfield  {title} {\enquote {\bibinfo {title} {Topological
  properties and dynamics of magnetic skyrmions},}\ }\href@noop {} {\bibfield
  {journal} {\bibinfo  {journal} {Nature Nanotechnology}\ }\textbf {\bibinfo
  {volume} {8}},\ \bibinfo {pages} {899--911} (\bibinfo {year}
  {2013})}\BibitemShut {NoStop}%
\bibitem [{\citenamefont
  {Dzyaloshinsky}(1958)}]{dzyaloshinsky1958thermodynamic}%
  \BibitemOpen
  \bibfield  {author} {\bibinfo {author} {\bibfnamefont {I.}~\bibnamefont
  {Dzyaloshinsky}},\ }\bibfield  {title} {\enquote {\bibinfo {title} {A
  thermodynamic theory of “weak” ferromagnetism of antiferromagnetics},}\
  }\href@noop {} {\bibfield  {journal} {\bibinfo  {journal} {Journal of Physics
  and Chemistry of Solids}\ }\textbf {\bibinfo {volume} {4}},\ \bibinfo {pages}
  {241--255} (\bibinfo {year} {1958})}\BibitemShut {NoStop}%
\bibitem [{\citenamefont {Moriya}(1960)}]{moriya1960anisotropic}%
  \BibitemOpen
  \bibfield  {author} {\bibinfo {author} {\bibfnamefont {T.}~\bibnamefont
  {Moriya}},\ }\bibfield  {title} {\enquote {\bibinfo {title} {Anisotropic
  superexchange interaction and weak ferromagnetism},}\ }\href@noop {}
  {\bibfield  {journal} {\bibinfo  {journal} {Physical Review}\ }\textbf
  {\bibinfo {volume} {120}},\ \bibinfo {pages} {91} (\bibinfo {year}
  {1960})}\BibitemShut {NoStop}%
\bibitem [{\citenamefont {Ojha}\ \emph {et~al.}(2021)\citenamefont {Ojha},
  \citenamefont {Mallick}, \citenamefont {Sharma}, \citenamefont {Thiaville},
  \citenamefont {Rohart},\ and\ \citenamefont {Bedanta}}]{ojha2021driving}%
  \BibitemOpen
  \bibfield  {author} {\bibinfo {author} {\bibfnamefont {B.}~\bibnamefont
  {Ojha}}, \bibinfo {author} {\bibfnamefont {S.}~\bibnamefont {Mallick}},
  \bibinfo {author} {\bibfnamefont {M.}~\bibnamefont {Sharma}}, \bibinfo
  {author} {\bibfnamefont {A.}~\bibnamefont {Thiaville}}, \bibinfo {author}
  {\bibfnamefont {S.}~\bibnamefont {Rohart}},\ and\ \bibinfo {author}
  {\bibfnamefont {S.}~\bibnamefont {Bedanta}},\ }\bibfield  {title} {\enquote
  {\bibinfo {title} {Driving skyrmions with low threshold current density in
  amorphous cofeb thin film},}\ }\href@noop {} {\bibfield  {journal} {\bibinfo
  {journal} {arXiv preprint arXiv:2106.02407}\ } (\bibinfo {year}
  {2021})}\BibitemShut {NoStop}%
\bibitem [{\citenamefont {Liu}\ \emph {et~al.}(2015{\natexlab{a}})\citenamefont
  {Liu}, \citenamefont {Zhu}, \citenamefont {Zhang}, \citenamefont {Liu},\ and\
  \citenamefont {Wang}}]{liu2015static}%
  \BibitemOpen
  \bibfield  {author} {\bibinfo {author} {\bibfnamefont {X.}~\bibnamefont
  {Liu}}, \bibinfo {author} {\bibfnamefont {Q.}~\bibnamefont {Zhu}}, \bibinfo
  {author} {\bibfnamefont {S.}~\bibnamefont {Zhang}}, \bibinfo {author}
  {\bibfnamefont {Q.}~\bibnamefont {Liu}},\ and\ \bibinfo {author}
  {\bibfnamefont {J.}~\bibnamefont {Wang}},\ }\bibfield  {title} {\enquote
  {\bibinfo {title} {Static property and current-driven precession of
  2$\pi$-vortex in nano-disk with dzyaloshinskii-moriya interaction},}\
  }\href@noop {} {\bibfield  {journal} {\bibinfo  {journal} {Aip Advances}\
  }\textbf {\bibinfo {volume} {5}},\ \bibinfo {pages} {087137} (\bibinfo {year}
  {2015}{\natexlab{a}})}\BibitemShut {NoStop}%
\bibitem [{\citenamefont {Deger}, \citenamefont {Yavuz},\ and\ \citenamefont
  {Yildiz}(2019)}]{spindeger2019current}%
  \BibitemOpen
  \bibfield  {author} {\bibinfo {author} {\bibfnamefont {C.}~\bibnamefont
  {Deger}}, \bibinfo {author} {\bibfnamefont {I.}~\bibnamefont {Yavuz}},\ and\
  \bibinfo {author} {\bibfnamefont {F.}~\bibnamefont {Yildiz}},\ }\bibfield
  {title} {\enquote {\bibinfo {title} {Current-driven coherent skyrmion
  generation},}\ }\href@noop {} {\bibfield  {journal} {\bibinfo  {journal}
  {Scientific Reports}\ }\textbf {\bibinfo {volume} {9}},\ \bibinfo {pages}
  {1--8} (\bibinfo {year} {2019})}\BibitemShut {NoStop}%
\bibitem [{\citenamefont {Lin}\ \emph {et~al.}(2013)\citenamefont {Lin},
  \citenamefont {Reichhardt}, \citenamefont {Batista},\ and\ \citenamefont
  {Saxena}}]{spinlin2013driven}%
  \BibitemOpen
  \bibfield  {author} {\bibinfo {author} {\bibfnamefont {S.-Z.}\ \bibnamefont
  {Lin}}, \bibinfo {author} {\bibfnamefont {C.}~\bibnamefont {Reichhardt}},
  \bibinfo {author} {\bibfnamefont {C.~D.}\ \bibnamefont {Batista}},\ and\
  \bibinfo {author} {\bibfnamefont {A.}~\bibnamefont {Saxena}},\ }\bibfield
  {title} {\enquote {\bibinfo {title} {Driven skyrmions and dynamical
  transitions in chiral magnets},}\ }\href@noop {} {\bibfield  {journal}
  {\bibinfo  {journal} {Physical Review Letters}\ }\textbf {\bibinfo {volume}
  {110}},\ \bibinfo {pages} {207202} (\bibinfo {year} {2013})}\BibitemShut
  {NoStop}%
\bibitem [{\citenamefont {Liu}\ \emph {et~al.}(2015{\natexlab{b}})\citenamefont
  {Liu}, \citenamefont {Du}, \citenamefont {Jia},\ and\ \citenamefont
  {Du}}]{spinliu2015switching}%
  \BibitemOpen
  \bibfield  {author} {\bibinfo {author} {\bibfnamefont {Y.}~\bibnamefont
  {Liu}}, \bibinfo {author} {\bibfnamefont {H.}~\bibnamefont {Du}}, \bibinfo
  {author} {\bibfnamefont {M.}~\bibnamefont {Jia}},\ and\ \bibinfo {author}
  {\bibfnamefont {A.}~\bibnamefont {Du}},\ }\bibfield  {title} {\enquote
  {\bibinfo {title} {Switching of a target skyrmion by a spin-polarized
  current},}\ }\href@noop {} {\bibfield  {journal} {\bibinfo  {journal}
  {Physical Review B}\ }\textbf {\bibinfo {volume} {91}},\ \bibinfo {pages}
  {094425} (\bibinfo {year} {2015}{\natexlab{b}})}\BibitemShut {NoStop}%
\bibitem [{\citenamefont {Moutafis}, \citenamefont {Komineas},\ and\
  \citenamefont {Bland}(2009)}]{magfielddynamics}%
  \BibitemOpen
  \bibfield  {author} {\bibinfo {author} {\bibfnamefont {C.}~\bibnamefont
  {Moutafis}}, \bibinfo {author} {\bibfnamefont {S.}~\bibnamefont {Komineas}},\
  and\ \bibinfo {author} {\bibfnamefont {J.~A.~C.}\ \bibnamefont {Bland}},\
  }\bibfield  {title} {\enquote {\bibinfo {title} {Dynamics and switching
  processes for magnetic bubbles in nanoelements},}\ }\href
  {https://doi.org/10.1103/PhysRevB.79.224429} {\bibfield  {journal} {\bibinfo
  {journal} {Phys. Rev. B}\ }\textbf {\bibinfo {volume} {79}},\ \bibinfo
  {pages} {224429} (\bibinfo {year} {2009})}\BibitemShut {NoStop}%
\bibitem [{\citenamefont {B{\"u}ttner}\ \emph {et~al.}(2015)\citenamefont
  {B{\"u}ttner}, \citenamefont {Moutafis}, \citenamefont {Schneider},
  \citenamefont {Kr{\"u}ger}, \citenamefont {G{\"u}nther}, \citenamefont
  {Geilhufe}, \citenamefont {Schmising}, \citenamefont {Mohanty}, \citenamefont
  {Pfau}, \citenamefont {Schaffert} \emph
  {et~al.}}]{magfieldbuttner2015dynamics}%
  \BibitemOpen
  \bibfield  {author} {\bibinfo {author} {\bibfnamefont {F.}~\bibnamefont
  {B{\"u}ttner}}, \bibinfo {author} {\bibfnamefont {C.}~\bibnamefont
  {Moutafis}}, \bibinfo {author} {\bibfnamefont {M.}~\bibnamefont {Schneider}},
  \bibinfo {author} {\bibfnamefont {B.}~\bibnamefont {Kr{\"u}ger}}, \bibinfo
  {author} {\bibfnamefont {C.}~\bibnamefont {G{\"u}nther}}, \bibinfo {author}
  {\bibfnamefont {J.}~\bibnamefont {Geilhufe}}, \bibinfo {author}
  {\bibfnamefont {C.~v.~K.}\ \bibnamefont {Schmising}}, \bibinfo {author}
  {\bibfnamefont {J.}~\bibnamefont {Mohanty}}, \bibinfo {author} {\bibfnamefont
  {B.}~\bibnamefont {Pfau}}, \bibinfo {author} {\bibfnamefont {S.}~\bibnamefont
  {Schaffert}}, \emph {et~al.},\ }\bibfield  {title} {\enquote {\bibinfo
  {title} {Dynamics and inertia of skyrmionic spin structures},}\ }\href@noop
  {} {\bibfield  {journal} {\bibinfo  {journal} {Nature Physics}\ }\textbf
  {\bibinfo {volume} {11}},\ \bibinfo {pages} {225--228} (\bibinfo {year}
  {2015})}\BibitemShut {NoStop}%
\bibitem [{\citenamefont {Tomasello}\ \emph {et~al.}(2018)\citenamefont
  {Tomasello}, \citenamefont {Komineas}, \citenamefont {Siracusano},
  \citenamefont {Carpentieri},\ and\ \citenamefont
  {Finocchio}}]{anisotropygradPhysRev}%
  \BibitemOpen
  \bibfield  {author} {\bibinfo {author} {\bibfnamefont {R.}~\bibnamefont
  {Tomasello}}, \bibinfo {author} {\bibfnamefont {S.}~\bibnamefont {Komineas}},
  \bibinfo {author} {\bibfnamefont {G.}~\bibnamefont {Siracusano}}, \bibinfo
  {author} {\bibfnamefont {M.}~\bibnamefont {Carpentieri}},\ and\ \bibinfo
  {author} {\bibfnamefont {G.}~\bibnamefont {Finocchio}},\ }\bibfield  {title}
  {\enquote {\bibinfo {title} {Chiral skyrmions in an anisotropy gradient},}\
  }\href {https://doi.org/10.1103/PhysRevB.98.024421} {\bibfield  {journal}
  {\bibinfo  {journal} {Phys. Rev. B}\ }\textbf {\bibinfo {volume} {98}},\
  \bibinfo {pages} {024421} (\bibinfo {year} {2018})}\BibitemShut {NoStop}%
\bibitem [{\citenamefont {Fert}, \citenamefont {Cros},\ and\ \citenamefont
  {Sampaio}(2013)}]{fert2013skyrmions}%
  \BibitemOpen
  \bibfield  {author} {\bibinfo {author} {\bibfnamefont {A.}~\bibnamefont
  {Fert}}, \bibinfo {author} {\bibfnamefont {V.}~\bibnamefont {Cros}},\ and\
  \bibinfo {author} {\bibfnamefont {J.}~\bibnamefont {Sampaio}},\ }\bibfield
  {title} {\enquote {\bibinfo {title} {Skyrmions on the track},}\ }\href@noop
  {} {\bibfield  {journal} {\bibinfo  {journal} {Nature Nanotechnology}\
  }\textbf {\bibinfo {volume} {8}},\ \bibinfo {pages} {152--156} (\bibinfo
  {year} {2013})}\BibitemShut {NoStop}%
\bibitem [{\citenamefont {{Schulz}}\ \emph {et~al.}(2012)\citenamefont
  {{Schulz}}, \citenamefont {{Ritz}}, \citenamefont {{Bauer}}, \citenamefont
  {{Halder}}, \citenamefont {{Wagner}}, \citenamefont {{Franz}}, \citenamefont
  {{Pfleiderer}}, \citenamefont {{Everschor}}, \citenamefont {{Garst}},\ and\
  \citenamefont {{Rosch}}}]{SkHE2012NatPh}%
  \BibitemOpen
  \bibfield  {author} {\bibinfo {author} {\bibfnamefont {T.}~\bibnamefont
  {{Schulz}}}, \bibinfo {author} {\bibfnamefont {R.}~\bibnamefont {{Ritz}}},
  \bibinfo {author} {\bibfnamefont {A.}~\bibnamefont {{Bauer}}}, \bibinfo
  {author} {\bibfnamefont {M.}~\bibnamefont {{Halder}}}, \bibinfo {author}
  {\bibfnamefont {M.}~\bibnamefont {{Wagner}}}, \bibinfo {author}
  {\bibfnamefont {C.}~\bibnamefont {{Franz}}}, \bibinfo {author} {\bibfnamefont
  {C.}~\bibnamefont {{Pfleiderer}}}, \bibinfo {author} {\bibfnamefont
  {K.}~\bibnamefont {{Everschor}}}, \bibinfo {author} {\bibfnamefont
  {M.}~\bibnamefont {{Garst}}},\ and\ \bibinfo {author} {\bibfnamefont
  {A.}~\bibnamefont {{Rosch}}},\ }\bibfield  {title} {\enquote {\bibinfo
  {title} {{Emergent electrodynamics of skyrmions in a chiral magnet}},}\
  }\href {https://doi.org/10.1038/nphys2231} {\bibfield  {journal} {\bibinfo
  {journal} {Nature Physics}\ }\textbf {\bibinfo {volume} {8}},\ \bibinfo
  {pages} {301--304} (\bibinfo {year} {2012})}\BibitemShut {NoStop}%
\bibitem [{\citenamefont {Jiang}\ \emph
  {et~al.}(2017{\natexlab{b}})\citenamefont {Jiang}, \citenamefont {Zhang},
  \citenamefont {Yu}, \citenamefont {Zhang}, \citenamefont {Wang},
  \citenamefont {Jungfleisch}, \citenamefont {Pearson}, \citenamefont {Cheng},
  \citenamefont {Heinonen}, \citenamefont {Wang} \emph
  {et~al.}}]{SkHEjiang2017direct}%
  \BibitemOpen
  \bibfield  {author} {\bibinfo {author} {\bibfnamefont {W.}~\bibnamefont
  {Jiang}}, \bibinfo {author} {\bibfnamefont {X.}~\bibnamefont {Zhang}},
  \bibinfo {author} {\bibfnamefont {G.}~\bibnamefont {Yu}}, \bibinfo {author}
  {\bibfnamefont {W.}~\bibnamefont {Zhang}}, \bibinfo {author} {\bibfnamefont
  {X.}~\bibnamefont {Wang}}, \bibinfo {author} {\bibfnamefont {M.~B.}\
  \bibnamefont {Jungfleisch}}, \bibinfo {author} {\bibfnamefont {J.~E.}\
  \bibnamefont {Pearson}}, \bibinfo {author} {\bibfnamefont {X.}~\bibnamefont
  {Cheng}}, \bibinfo {author} {\bibfnamefont {O.}~\bibnamefont {Heinonen}},
  \bibinfo {author} {\bibfnamefont {K.~L.}\ \bibnamefont {Wang}}, \emph
  {et~al.},\ }\bibfield  {title} {\enquote {\bibinfo {title} {Direct
  observation of the skyrmion hall effect},}\ }\href@noop {} {\bibfield
  {journal} {\bibinfo  {journal} {Nature Physics}\ }\textbf {\bibinfo {volume}
  {13}},\ \bibinfo {pages} {162--169} (\bibinfo {year}
  {2017}{\natexlab{b}})}\BibitemShut {NoStop}%
\bibitem [{\citenamefont {Litzius}\ \emph {et~al.}(2017)\citenamefont
  {Litzius}, \citenamefont {Lemesh}, \citenamefont {Kr{\"u}ger}, \citenamefont
  {Bassirian}, \citenamefont {Caretta}, \citenamefont {Richter}, \citenamefont
  {B{\"u}ttner}, \citenamefont {Sato}, \citenamefont {Tretiakov}, \citenamefont
  {F{\"o}rster} \emph {et~al.}}]{SkHElitzius2017skyrmion}%
  \BibitemOpen
  \bibfield  {author} {\bibinfo {author} {\bibfnamefont {K.}~\bibnamefont
  {Litzius}}, \bibinfo {author} {\bibfnamefont {I.}~\bibnamefont {Lemesh}},
  \bibinfo {author} {\bibfnamefont {B.}~\bibnamefont {Kr{\"u}ger}}, \bibinfo
  {author} {\bibfnamefont {P.}~\bibnamefont {Bassirian}}, \bibinfo {author}
  {\bibfnamefont {L.}~\bibnamefont {Caretta}}, \bibinfo {author} {\bibfnamefont
  {K.}~\bibnamefont {Richter}}, \bibinfo {author} {\bibfnamefont
  {F.}~\bibnamefont {B{\"u}ttner}}, \bibinfo {author} {\bibfnamefont
  {K.}~\bibnamefont {Sato}}, \bibinfo {author} {\bibfnamefont {O.~A.}\
  \bibnamefont {Tretiakov}}, \bibinfo {author} {\bibfnamefont {J.}~\bibnamefont
  {F{\"o}rster}}, \emph {et~al.},\ }\bibfield  {title} {\enquote {\bibinfo
  {title} {Skyrmion hall effect revealed by direct time-resolved x-ray
  microscopy},}\ }\href@noop {} {\bibfield  {journal} {\bibinfo  {journal}
  {Nature Physics}\ }\textbf {\bibinfo {volume} {13}},\ \bibinfo {pages}
  {170--175} (\bibinfo {year} {2017})}\BibitemShut {NoStop}%
\bibitem [{\citenamefont {Zhang}\ \emph {et~al.}(2017)\citenamefont {Zhang},
  \citenamefont {Luo}, \citenamefont {Yan}, \citenamefont {Ou-Yang},
  \citenamefont {Yang}, \citenamefont {Chen}, \citenamefont {Zhu},\ and\
  \citenamefont {You}}]{overcomeSkhezhang2017magnetic}%
  \BibitemOpen
  \bibfield  {author} {\bibinfo {author} {\bibfnamefont {Y.}~\bibnamefont
  {Zhang}}, \bibinfo {author} {\bibfnamefont {S.}~\bibnamefont {Luo}}, \bibinfo
  {author} {\bibfnamefont {B.}~\bibnamefont {Yan}}, \bibinfo {author}
  {\bibfnamefont {J.}~\bibnamefont {Ou-Yang}}, \bibinfo {author} {\bibfnamefont
  {X.}~\bibnamefont {Yang}}, \bibinfo {author} {\bibfnamefont {S.}~\bibnamefont
  {Chen}}, \bibinfo {author} {\bibfnamefont {B.}~\bibnamefont {Zhu}},\ and\
  \bibinfo {author} {\bibfnamefont {L.}~\bibnamefont {You}},\ }\bibfield
  {title} {\enquote {\bibinfo {title} {Magnetic skyrmions without the skyrmion
  hall effect in a magnetic nanotrack with perpendicular anisotropy},}\
  }\href@noop {} {\bibfield  {journal} {\bibinfo  {journal} {Nanoscale}\
  }\textbf {\bibinfo {volume} {9}},\ \bibinfo {pages} {10212--10218} (\bibinfo
  {year} {2017})}\BibitemShut {NoStop}%
\bibitem [{\citenamefont {Toscano}\ \emph {et~al.}(2020)\citenamefont
  {Toscano}, \citenamefont {Mendon{\c{c}}a}, \citenamefont {Miranda},
  \citenamefont {de~Araujo}, \citenamefont {Sato}, \citenamefont {Coura},\ and\
  \citenamefont {Leonel}}]{toscano2020suppression}%
  \BibitemOpen
  \bibfield  {author} {\bibinfo {author} {\bibfnamefont {D.}~\bibnamefont
  {Toscano}}, \bibinfo {author} {\bibfnamefont {J.}~\bibnamefont
  {Mendon{\c{c}}a}}, \bibinfo {author} {\bibfnamefont {A.}~\bibnamefont
  {Miranda}}, \bibinfo {author} {\bibfnamefont {C.}~\bibnamefont {de~Araujo}},
  \bibinfo {author} {\bibfnamefont {F.}~\bibnamefont {Sato}}, \bibinfo {author}
  {\bibfnamefont {P.}~\bibnamefont {Coura}},\ and\ \bibinfo {author}
  {\bibfnamefont {S.}~\bibnamefont {Leonel}},\ }\bibfield  {title} {\enquote
  {\bibinfo {title} {Suppression of the skyrmion hall effect in planar
  nanomagnets by the magnetic properties engineering: Skyrmion transport on
  nanotracks with magnetic strips},}\ }\href@noop {} {\bibfield  {journal}
  {\bibinfo  {journal} {Journal of Magnetism and Magnetic Materials}\ }\textbf
  {\bibinfo {volume} {504}},\ \bibinfo {pages} {166655} (\bibinfo {year}
  {2020})}\BibitemShut {NoStop}%
\bibitem [{\citenamefont {Zhang}, \citenamefont {Zhou},\ and\ \citenamefont
  {Ezawa}(2016)}]{safzhang2016magnetic}%
  \BibitemOpen
  \bibfield  {author} {\bibinfo {author} {\bibfnamefont {X.}~\bibnamefont
  {Zhang}}, \bibinfo {author} {\bibfnamefont {Y.}~\bibnamefont {Zhou}},\ and\
  \bibinfo {author} {\bibfnamefont {M.}~\bibnamefont {Ezawa}},\ }\bibfield
  {title} {\enquote {\bibinfo {title} {Magnetic bilayer-skyrmions without
  skyrmion hall effect},}\ }\href@noop {} {\bibfield  {journal} {\bibinfo
  {journal} {Nature Communications}\ }\textbf {\bibinfo {volume} {7}},\
  \bibinfo {pages} {1--7} (\bibinfo {year} {2016})}\BibitemShut {NoStop}%
\bibitem [{\citenamefont {Zhang}, \citenamefont {Ezawa},\ and\ \citenamefont
  {Zhou}(2016)}]{safzhang2016thermally}%
  \BibitemOpen
  \bibfield  {author} {\bibinfo {author} {\bibfnamefont {X.}~\bibnamefont
  {Zhang}}, \bibinfo {author} {\bibfnamefont {M.}~\bibnamefont {Ezawa}},\ and\
  \bibinfo {author} {\bibfnamefont {Y.}~\bibnamefont {Zhou}},\ }\bibfield
  {title} {\enquote {\bibinfo {title} {Thermally stable magnetic skyrmions in
  multilayer synthetic antiferromagnetic racetracks},}\ }\href@noop {}
  {\bibfield  {journal} {\bibinfo  {journal} {Physical Review B}\ }\textbf
  {\bibinfo {volume} {94}},\ \bibinfo {pages} {064406} (\bibinfo {year}
  {2016})}\BibitemShut {NoStop}%
\bibitem [{\citenamefont {Beg}\ \emph {et~al.}(2015)\citenamefont {Beg},
  \citenamefont {Carey}, \citenamefont {Wang}, \citenamefont
  {Cort{\'e}s-Ortu{\~n}o}, \citenamefont {Vousden}, \citenamefont {Bisotti},
  \citenamefont {Albert}, \citenamefont {Chernyshenko}, \citenamefont
  {Hovorka}, \citenamefont {Stamps} \emph {et~al.}}]{SkrMbeg2015ground}%
  \BibitemOpen
  \bibfield  {author} {\bibinfo {author} {\bibfnamefont {M.}~\bibnamefont
  {Beg}}, \bibinfo {author} {\bibfnamefont {R.}~\bibnamefont {Carey}}, \bibinfo
  {author} {\bibfnamefont {W.}~\bibnamefont {Wang}}, \bibinfo {author}
  {\bibfnamefont {D.}~\bibnamefont {Cort{\'e}s-Ortu{\~n}o}}, \bibinfo {author}
  {\bibfnamefont {M.}~\bibnamefont {Vousden}}, \bibinfo {author} {\bibfnamefont
  {M.-A.}\ \bibnamefont {Bisotti}}, \bibinfo {author} {\bibfnamefont
  {M.}~\bibnamefont {Albert}}, \bibinfo {author} {\bibfnamefont
  {D.}~\bibnamefont {Chernyshenko}}, \bibinfo {author} {\bibfnamefont
  {O.}~\bibnamefont {Hovorka}}, \bibinfo {author} {\bibfnamefont {R.~L.}\
  \bibnamefont {Stamps}}, \emph {et~al.},\ }\bibfield  {title} {\enquote
  {\bibinfo {title} {Ground state search, hysteretic behaviour and reversal
  mechanism of skyrmionic textures in confined helimagnetic nanostructures},}\
  }\href@noop {} {\bibfield  {journal} {\bibinfo  {journal} {Scientific
  Reports}\ }\textbf {\bibinfo {volume} {5}},\ \bibinfo {pages} {1--14}
  (\bibinfo {year} {2015})}\BibitemShut {NoStop}%
\bibitem [{\citenamefont {Rohart}\ and\ \citenamefont
  {Thiaville}(2013)}]{SkrMrohart2013skyrmion}%
  \BibitemOpen
  \bibfield  {author} {\bibinfo {author} {\bibfnamefont {S.}~\bibnamefont
  {Rohart}}\ and\ \bibinfo {author} {\bibfnamefont {A.}~\bibnamefont
  {Thiaville}},\ }\bibfield  {title} {\enquote {\bibinfo {title} {Skyrmion
  confinement in ultrathin film nanostructures in the presence of
  dzyaloshinskii-moriya interaction},}\ }\href@noop {} {\bibfield  {journal}
  {\bibinfo  {journal} {Physical Review B}\ }\textbf {\bibinfo {volume} {88}},\
  \bibinfo {pages} {184422} (\bibinfo {year} {2013})}\BibitemShut {NoStop}%
\bibitem [{\citenamefont {Bogdanov}\ and\ \citenamefont
  {Hubert}(1999)}]{skyrMbogdanov1999stability}%
  \BibitemOpen
  \bibfield  {author} {\bibinfo {author} {\bibfnamefont {A.}~\bibnamefont
  {Bogdanov}}\ and\ \bibinfo {author} {\bibfnamefont {A.}~\bibnamefont
  {Hubert}},\ }\bibfield  {title} {\enquote {\bibinfo {title} {The stability of
  vortex-like structures in uniaxial ferromagnets},}\ }\href@noop {} {\bibfield
   {journal} {\bibinfo  {journal} {Journal of Magnetism and Magnetic
  materials}\ }\textbf {\bibinfo {volume} {195}},\ \bibinfo {pages} {182--192}
  (\bibinfo {year} {1999})}\BibitemShut {NoStop}%
\bibitem [{\citenamefont {Zhang}\ \emph {et~al.}(2020)\citenamefont {Zhang},
  \citenamefont {Zhu}, \citenamefont {Kang}, \citenamefont {Zhang},\ and\
  \citenamefont {Zhao}}]{after27para3PhysRevApplied.13.054049}%
  \BibitemOpen
  \bibfield  {author} {\bibinfo {author} {\bibfnamefont {H.}~\bibnamefont
  {Zhang}}, \bibinfo {author} {\bibfnamefont {D.}~\bibnamefont {Zhu}}, \bibinfo
  {author} {\bibfnamefont {W.}~\bibnamefont {Kang}}, \bibinfo {author}
  {\bibfnamefont {Y.}~\bibnamefont {Zhang}},\ and\ \bibinfo {author}
  {\bibfnamefont {W.}~\bibnamefont {Zhao}},\ }\bibfield  {title} {\enquote
  {\bibinfo {title} {Stochastic computing implemented by skyrmionic logic
  devices},}\ }\href {https://doi.org/10.1103/PhysRevApplied.13.054049}
  {\bibfield  {journal} {\bibinfo  {journal} {Phys. Rev. Applied}\ }\textbf
  {\bibinfo {volume} {13}},\ \bibinfo {pages} {054049} (\bibinfo {year}
  {2020})}\BibitemShut {NoStop}%
\bibitem [{\citenamefont {Behera}\ \emph {et~al.}(2020)\citenamefont {Behera},
  \citenamefont {Murapaka}, \citenamefont {Mallick}, \citenamefont {Singh},\
  and\ \citenamefont {Bedanta}}]{behera2020skyrmion}%
  \BibitemOpen
  \bibfield  {author} {\bibinfo {author} {\bibfnamefont {A.~K.}\ \bibnamefont
  {Behera}}, \bibinfo {author} {\bibfnamefont {C.}~\bibnamefont {Murapaka}},
  \bibinfo {author} {\bibfnamefont {S.}~\bibnamefont {Mallick}}, \bibinfo
  {author} {\bibfnamefont {B.~B.}\ \bibnamefont {Singh}},\ and\ \bibinfo
  {author} {\bibfnamefont {S.}~\bibnamefont {Bedanta}},\ }\bibfield  {title}
  {\enquote {\bibinfo {title} {Skyrmion racetrack memory with an antidot},}\
  }\href@noop {} {\bibfield  {journal} {\bibinfo  {journal} {Journal of Physics
  D: Applied Physics}\ }\textbf {\bibinfo {volume} {54}},\ \bibinfo {pages}
  {025001} (\bibinfo {year} {2020})}\BibitemShut {NoStop}%
\bibitem [{\citenamefont {Chen}\ \emph {et~al.}(2017)\citenamefont {Chen},
  \citenamefont {Kang}, \citenamefont {Zhu}, \citenamefont {Zhang},
  \citenamefont {Lei}, \citenamefont {Zhang}, \citenamefont {Zhou},\ and\
  \citenamefont {ZHAO}}]{para3aft1article}%
  \BibitemOpen
  \bibfield  {author} {\bibinfo {author} {\bibfnamefont {X.}~\bibnamefont
  {Chen}}, \bibinfo {author} {\bibfnamefont {W.}~\bibnamefont {Kang}}, \bibinfo
  {author} {\bibfnamefont {D.}~\bibnamefont {Zhu}}, \bibinfo {author}
  {\bibfnamefont {X.}~\bibnamefont {Zhang}}, \bibinfo {author} {\bibfnamefont
  {n.}~\bibnamefont {Lei}}, \bibinfo {author} {\bibfnamefont {Y.}~\bibnamefont
  {Zhang}}, \bibinfo {author} {\bibfnamefont {Y.}~\bibnamefont {Zhou}},\ and\
  \bibinfo {author} {\bibfnamefont {W.}~\bibnamefont {ZHAO}},\ }\bibfield
  {title} {\enquote {\bibinfo {title} {Skyrmion dynamics in width-varying
  nanotracks and implications for skyrmionic applications},}\ }\href
  {https://doi.org/10.1063/1.5005953} {\bibfield  {journal} {\bibinfo
  {journal} {Applied Physics Letters}\ }\textbf {\bibinfo {volume} {111}},\
  \bibinfo {pages} {202406} (\bibinfo {year} {2017})}\BibitemShut {NoStop}%
\bibitem [{\citenamefont {Liu}\ \emph {et~al.}(2017)\citenamefont {Liu},
  \citenamefont {Lei}, \citenamefont {Zhao}, \citenamefont {Liu}, \citenamefont
  {Ruotolo}, \citenamefont {Braun},\ and\ \citenamefont {Zhou}}]{para3aft2}%
  \BibitemOpen
  \bibfield  {author} {\bibinfo {author} {\bibfnamefont {Y.}~\bibnamefont
  {Liu}}, \bibinfo {author} {\bibfnamefont {N.}~\bibnamefont {Lei}}, \bibinfo
  {author} {\bibfnamefont {W.}~\bibnamefont {Zhao}}, \bibinfo {author}
  {\bibfnamefont {W.}~\bibnamefont {Liu}}, \bibinfo {author} {\bibfnamefont
  {A.}~\bibnamefont {Ruotolo}}, \bibinfo {author} {\bibfnamefont {H.-B.}\
  \bibnamefont {Braun}},\ and\ \bibinfo {author} {\bibfnamefont
  {Y.}~\bibnamefont {Zhou}},\ }\bibfield  {title} {\enquote {\bibinfo {title}
  {Chopping skyrmions from magnetic chiral domains with uniaxial stress in
  magnetic nanowire},}\ }\href {https://doi.org/10.1063/1.4993433} {\bibfield
  {journal} {\bibinfo  {journal} {Applied Physics Letters}\ }\textbf {\bibinfo
  {volume} {111}},\ \bibinfo {pages} {022406} (\bibinfo {year}
  {2017})}\BibitemShut {NoStop}%
\bibitem [{\citenamefont {Zhang}\ \emph
  {et~al.}(2015{\natexlab{b}})\citenamefont {Zhang}, \citenamefont {Zhao},
  \citenamefont {Fangohr}, \citenamefont {Liu}, \citenamefont {Xia},
  \citenamefont {Xia},\ and\ \citenamefont {Morvan}}]{zhang2015skyrmion}%
  \BibitemOpen
  \bibfield  {author} {\bibinfo {author} {\bibfnamefont {X.}~\bibnamefont
  {Zhang}}, \bibinfo {author} {\bibfnamefont {G.}~\bibnamefont {Zhao}},
  \bibinfo {author} {\bibfnamefont {H.}~\bibnamefont {Fangohr}}, \bibinfo
  {author} {\bibfnamefont {J.~P.}\ \bibnamefont {Liu}}, \bibinfo {author}
  {\bibfnamefont {W.}~\bibnamefont {Xia}}, \bibinfo {author} {\bibfnamefont
  {J.}~\bibnamefont {Xia}},\ and\ \bibinfo {author} {\bibfnamefont
  {F.}~\bibnamefont {Morvan}},\ }\bibfield  {title} {\enquote {\bibinfo {title}
  {Skyrmion-skyrmion and skyrmion-edge repulsions in skyrmion-based racetrack
  memory},}\ }\href@noop {} {\bibfield  {journal} {\bibinfo  {journal}
  {Scientific Reports}\ }\textbf {\bibinfo {volume} {5}},\ \bibinfo {pages}
  {1--6} (\bibinfo {year} {2015}{\natexlab{b}})}\BibitemShut {NoStop}%
\bibitem [{\citenamefont {Zhang}\ \emph {et~al.}(2016)\citenamefont {Zhang},
  \citenamefont {Xia}, \citenamefont {Zhou}, \citenamefont {Wang},
  \citenamefont {Liu}, \citenamefont {Zhao},\ and\ \citenamefont
  {Ezawa}}]{motivationpaper}%
  \BibitemOpen
  \bibfield  {author} {\bibinfo {author} {\bibfnamefont {X.}~\bibnamefont
  {Zhang}}, \bibinfo {author} {\bibfnamefont {J.}~\bibnamefont {Xia}}, \bibinfo
  {author} {\bibfnamefont {Y.}~\bibnamefont {Zhou}}, \bibinfo {author}
  {\bibfnamefont {D.}~\bibnamefont {Wang}}, \bibinfo {author} {\bibfnamefont
  {X.}~\bibnamefont {Liu}}, \bibinfo {author} {\bibfnamefont {W.}~\bibnamefont
  {Zhao}},\ and\ \bibinfo {author} {\bibfnamefont {M.}~\bibnamefont {Ezawa}},\
  }\bibfield  {title} {\enquote {\bibinfo {title} {Control and manipulation of
  a magnetic skyrmionium in nanostructures},}\ }\href
  {https://doi.org/10.1103/PhysRevB.94.094420} {\bibfield  {journal} {\bibinfo
  {journal} {Physical Review B}\ }\textbf {\bibinfo {volume} {94}},\ \bibinfo
  {pages} {094420} (\bibinfo {year} {2016})}\BibitemShut {NoStop}%
\bibitem [{\citenamefont {Iwasaki}, \citenamefont {Mochizuki},\ and\
  \citenamefont {Nagaosa}(2013)}]{iwasaki2013current}%
  \BibitemOpen
  \bibfield  {author} {\bibinfo {author} {\bibfnamefont {J.}~\bibnamefont
  {Iwasaki}}, \bibinfo {author} {\bibfnamefont {M.}~\bibnamefont {Mochizuki}},\
  and\ \bibinfo {author} {\bibfnamefont {N.}~\bibnamefont {Nagaosa}},\
  }\bibfield  {title} {\enquote {\bibinfo {title} {Current-induced skyrmion
  dynamics in constricted geometries},}\ }\href@noop {} {\bibfield  {journal}
  {\bibinfo  {journal} {Nature nanotechnology}\ }\textbf {\bibinfo {volume}
  {8}},\ \bibinfo {pages} {742--747} (\bibinfo {year} {2013})}\BibitemShut
  {NoStop}%
\bibitem [{\citenamefont {Zhou}\ and\ \citenamefont
  {Ezawa}(2014)}]{HYBRIDDEVICEzhou2014reversible}%
  \BibitemOpen
  \bibfield  {author} {\bibinfo {author} {\bibfnamefont {Y.}~\bibnamefont
  {Zhou}}\ and\ \bibinfo {author} {\bibfnamefont {M.}~\bibnamefont {Ezawa}},\
  }\bibfield  {title} {\enquote {\bibinfo {title} {A reversible conversion
  between a skyrmion and a domain-wall pair in a junction geometry},}\
  }\href@noop {} {\bibfield  {journal} {\bibinfo  {journal} {Nature
  Communications}\ }\textbf {\bibinfo {volume} {5}},\ \bibinfo {pages} {1--8}
  (\bibinfo {year} {2014})}\BibitemShut {NoStop}%
\bibitem [{\citenamefont {Thomas}\ \emph {et~al.}(2012)\citenamefont {Thomas},
  \citenamefont {Hayashi}, \citenamefont {Moriya}, \citenamefont {Rettner},\
  and\ \citenamefont {Parkin}}]{thomas2012topological}%
  \BibitemOpen
  \bibfield  {author} {\bibinfo {author} {\bibfnamefont {L.}~\bibnamefont
  {Thomas}}, \bibinfo {author} {\bibfnamefont {M.}~\bibnamefont {Hayashi}},
  \bibinfo {author} {\bibfnamefont {R.}~\bibnamefont {Moriya}}, \bibinfo
  {author} {\bibfnamefont {C.}~\bibnamefont {Rettner}},\ and\ \bibinfo {author}
  {\bibfnamefont {S.}~\bibnamefont {Parkin}},\ }\bibfield  {title} {\enquote
  {\bibinfo {title} {Topological repulsion between domain walls in magnetic
  nanowires leading to the formation of bound states},}\ }\href@noop {}
  {\bibfield  {journal} {\bibinfo  {journal} {Nature communications}\ }\textbf
  {\bibinfo {volume} {3}},\ \bibinfo {pages} {1--7} (\bibinfo {year}
  {2012})}\BibitemShut {NoStop}%
\bibitem [{\citenamefont {P.~del Real}\ \emph {et~al.}(2017)\citenamefont
  {P.~del Real}, \citenamefont {Raposo}, \citenamefont {Martinez},\ and\
  \citenamefont {Hayashi}}]{p2017current}%
  \BibitemOpen
  \bibfield  {author} {\bibinfo {author} {\bibfnamefont {R.}~\bibnamefont
  {P.~del Real}}, \bibinfo {author} {\bibfnamefont {V.}~\bibnamefont {Raposo}},
  \bibinfo {author} {\bibfnamefont {E.}~\bibnamefont {Martinez}},\ and\
  \bibinfo {author} {\bibfnamefont {M.}~\bibnamefont {Hayashi}},\ }\bibfield
  {title} {\enquote {\bibinfo {title} {Current-induced generation and
  synchronous motion of highly packed coupled chiral domain walls},}\
  }\href@noop {} {\bibfield  {journal} {\bibinfo  {journal} {Nano letters}\
  }\textbf {\bibinfo {volume} {17}},\ \bibinfo {pages} {1814--1818} (\bibinfo
  {year} {2017})}\BibitemShut {NoStop}%
\bibitem [{\citenamefont {Krishnia}\ \emph {et~al.}(2017)\citenamefont
  {Krishnia}, \citenamefont {Sethi}, \citenamefont {Gan}, \citenamefont
  {Kholid}, \citenamefont {Purnama}, \citenamefont {Ramu}, \citenamefont
  {Herng}, \citenamefont {Ding},\ and\ \citenamefont {Lew}}]{RKKYtorquedoi}%
  \BibitemOpen
  \bibfield  {author} {\bibinfo {author} {\bibfnamefont {S.}~\bibnamefont
  {Krishnia}}, \bibinfo {author} {\bibfnamefont {P.}~\bibnamefont {Sethi}},
  \bibinfo {author} {\bibfnamefont {W.}~\bibnamefont {Gan}}, \bibinfo {author}
  {\bibfnamefont {F.}~\bibnamefont {Kholid}}, \bibinfo {author} {\bibfnamefont
  {I.}~\bibnamefont {Purnama}}, \bibinfo {author} {\bibfnamefont
  {M.}~\bibnamefont {Ramu}}, \bibinfo {author} {\bibfnamefont {T.}~\bibnamefont
  {Herng}}, \bibinfo {author} {\bibfnamefont {J.}~\bibnamefont {Ding}},\ and\
  \bibinfo {author} {\bibfnamefont {W.}~\bibnamefont {Lew}},\ }\bibfield
  {title} {\enquote {\bibinfo {title} {Role of rkky torque on domain wall
  motion in synthetic antiferromagnetic nanowires with opposite spin hall
  angles},}\ }\href@noop {} {\bibfield  {journal} {\bibinfo  {journal}
  {Scientific Reports}\ }\textbf {\bibinfo {volume} {7}},\ \bibinfo {pages}
  {1--11} (\bibinfo {year} {2017})}\BibitemShut {NoStop}%
\bibitem [{\citenamefont {Grollier}\ \emph {et~al.}(2020)\citenamefont
  {Grollier}, \citenamefont {Querlioz}, \citenamefont {Camsari}, \citenamefont
  {Everschor-Sitte}, \citenamefont {Fukami},\ and\ \citenamefont
  {Stiles}}]{grollier2020neuromorphic}%
  \BibitemOpen
  \bibfield  {author} {\bibinfo {author} {\bibfnamefont {J.}~\bibnamefont
  {Grollier}}, \bibinfo {author} {\bibfnamefont {D.}~\bibnamefont {Querlioz}},
  \bibinfo {author} {\bibfnamefont {K.}~\bibnamefont {Camsari}}, \bibinfo
  {author} {\bibfnamefont {K.}~\bibnamefont {Everschor-Sitte}}, \bibinfo
  {author} {\bibfnamefont {S.}~\bibnamefont {Fukami}},\ and\ \bibinfo {author}
  {\bibfnamefont {M.~D.}\ \bibnamefont {Stiles}},\ }\bibfield  {title}
  {\enquote {\bibinfo {title} {Neuromorphic spintronics},}\ }\href@noop {}
  {\bibfield  {journal} {\bibinfo  {journal} {Nature Electronics}\ }\textbf
  {\bibinfo {volume} {3}},\ \bibinfo {pages} {360--370} (\bibinfo {year}
  {2020})}\BibitemShut {NoStop}%
\end{thebibliography}%


\begin{thebibliography}{0}%
\makeatletter
\providecommand \@ifxundefined [1]{%
 \@ifx{#1\undefined}
}%
\providecommand \@ifnum [1]{%
 \ifnum #1\expandafter \@firstoftwo
 \else \expandafter \@secondoftwo
 \fi
}%
\providecommand \@ifx [1]{%
 \ifx #1\expandafter \@firstoftwo
 \else \expandafter \@secondoftwo
 \fi
}%
\providecommand \natexlab [1]{#1}%
\providecommand \enquote  [1]{``#1''}%
\providecommand \bibnamefont  [1]{#1}%
\providecommand \bibfnamefont [1]{#1}%
\providecommand \citenamefont [1]{#1}%
\providecommand \href@noop [0]{\@secondoftwo}%
\providecommand \href [0]{\begingroup \@sanitize@url \@href}%
\providecommand \@href[1]{\@@startlink{#1}\@@href}%
\providecommand \@@href[1]{\endgroup#1\@@endlink}%
\providecommand \@sanitize@url [0]{\catcode `\\12\catcode `\$12\catcode
  `\&12\catcode `\#12\catcode `\^12\catcode `\_12\catcode `\%12\relax}%
\providecommand \@@startlink[1]{}%
\providecommand \@@endlink[0]{}%
\providecommand \url  [0]{\begingroup\@sanitize@url \@url }%
\providecommand \@url [1]{\endgroup\@href {#1}{\urlprefix }}%
\providecommand \urlprefix  [0]{URL }%
\providecommand \Eprint [0]{\href }%
\providecommand \doibase [0]{https://doi.org/}%
\providecommand \selectlanguage [0]{\@gobble}%
\providecommand \bibinfo  [0]{\@secondoftwo}%
\providecommand \bibfield  [0]{\@secondoftwo}%
\providecommand \translation [1]{[#1]}%
\providecommand \BibitemOpen [0]{}%
\providecommand \bibitemStop [0]{}%
\providecommand \bibitemNoStop [0]{.\EOS\space}%
\providecommand \EOS [0]{\spacefactor3000\relax}%
\providecommand \BibitemShut  [1]{\csname bibitem#1\endcsname}%
\let\auto@bib@innerbib\@empty
\end{thebibliography}%

\end{document}


\preprint{ Brindaban {\it et al.}, Driving skyrmions with low threshold current density in amorphous CoFeB thin film}

\title[Title]{Device geometry dependent deterministic skyrmion generation from a skyrmionium}

\author{Adyashakti Dash}
\address {Laboratory for Nanomagnetism and Magnetic Materials, School of Physical Sciences, National Institute of Science Education and Research (NISER), HBNI, Jatni 752050, India}
 
\author{Brindaban Ojha}
\address{Laboratory for Nanomagnetism and Magnetic Materials, School of Physical Sciences, National Institute of Science Education and Research (NISER), HBNI, Jatni 752050, India}

\author{Shaktiranjan Mohanty}
\address{Laboratory for Nanomagnetism and Magnetic Materials, School of Physical Sciences, National Institute of Science Education and Research (NISER), HBNI, Jatni 752050, India}

\author{Ashish Kumar Moharana}
\address{Laboratory for Nanomagnetism and Magnetic Materials, School of Physical Sciences, National Institute of Science Education and Research (NISER), HBNI, Jatni 752050, India}

\author{Subhankar Bedanta$^*$}
\email{sbedanta@niser.ac.in}
\address{Laboratory for Nanomagnetism and Magnetic Materials, School of Physical Sciences, National Institute of Science Education and Research (NISER), HBNI, Jatni 752050, India}

\date{\today}


\pacs{}
\maketitle
\onecolumngrid

Fig. S1 shows the detailed design of Dvc-2. The divider junction-1 splits the domain wall and also provides a resistance in translation of spin texture. The junctions 1, 2 and 3 are potential pinning sites for DWs. The geometric asymmetry can be seen at notch which influences the skyrmions generation in single FM layer. This results in generation of different skyrmionic states in single FM layer. (gif)

\begin{figure}[h!]
	\centering
	\includegraphics[width=0.9\linewidth]{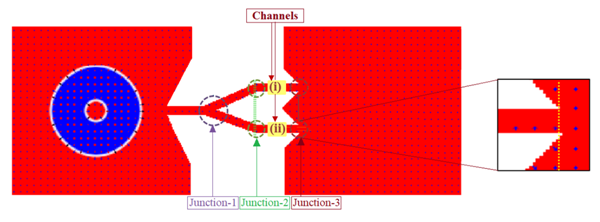}
	\renewcommand{\thefigure}{S\arabic{figure}}
	\caption{ Schematic of detailed design of Dvc-2 showing junctions and channels.}
\end{figure}

\leavevmode\thispagestyle{empty}\newpage

The formation of 3 skyrmions (generated from the first batch of DW pair) in Dvc-3 in SAF is illustrated in fig. S2. The conversion process starts at 0.17 ns when the skyrmionium touches the interface of narrow and wide channel. At 0.30 ns, there is injection of the former DW pair in the narrow channel. It translates until it experiences repulsion from the junction divider. The later DW pair is injected at 0.54 ns. The later DW pair is injected at 0.54 ns. The former DW pairs experience STT, repulsion from the later batch of DW pair, as well as RKKY torque. Hence, they gain enough energy to overcome the constriction barrier at the notches. The channel (iii) is slightly lengthier than the channel (i) which explains the delay in translation of DWs and release of skyrmions. The DW pair in channel (ii) travels the shortest distance and hence, it is released earlier than the rest. At current density, $J = 4.5 \times 10^{12} A/m^2$, the former domain wall pairs are released to generate skyrmions, however, the later domain wall pairs are pinned due to insufficient energy.  

\begin{figure}[h!]
	\centering
	\includegraphics[width=0.8\linewidth]{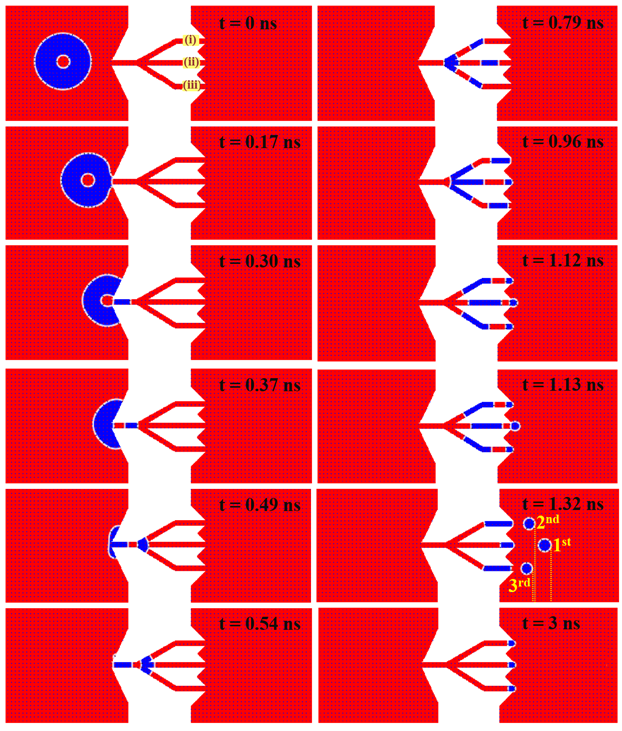}
	\renewcommand{\thefigure}{S\arabic{figure}}
	\caption{ Time evolution of spin textures during the conversion process in 1-3 channel design: Dvc-3 for SAF at $J = 4.5 \times 10^{12} A/m^2$.}
\end{figure}

\leavevmode\thispagestyle{empty}\newpage

Fig. S3 represents the generation of number of skyrmions at different $t$ and $J$ in single FM layer and SAF for all the devices, respectively. In single FM layer, all possible states have been observed for all three devices. But, in SAF, we get number of skyrmions equal to the number of channels or twice of it. The delay in release between skyrmions decrease as the STT in the system is increased. However, this can be deliberately tempered by adjusting the separating distance between the former and later batch of domains, as well as changing the length and width of the narrow channels.

\begin{figure}[h!]
	\centering
	\includegraphics[width=1\linewidth]{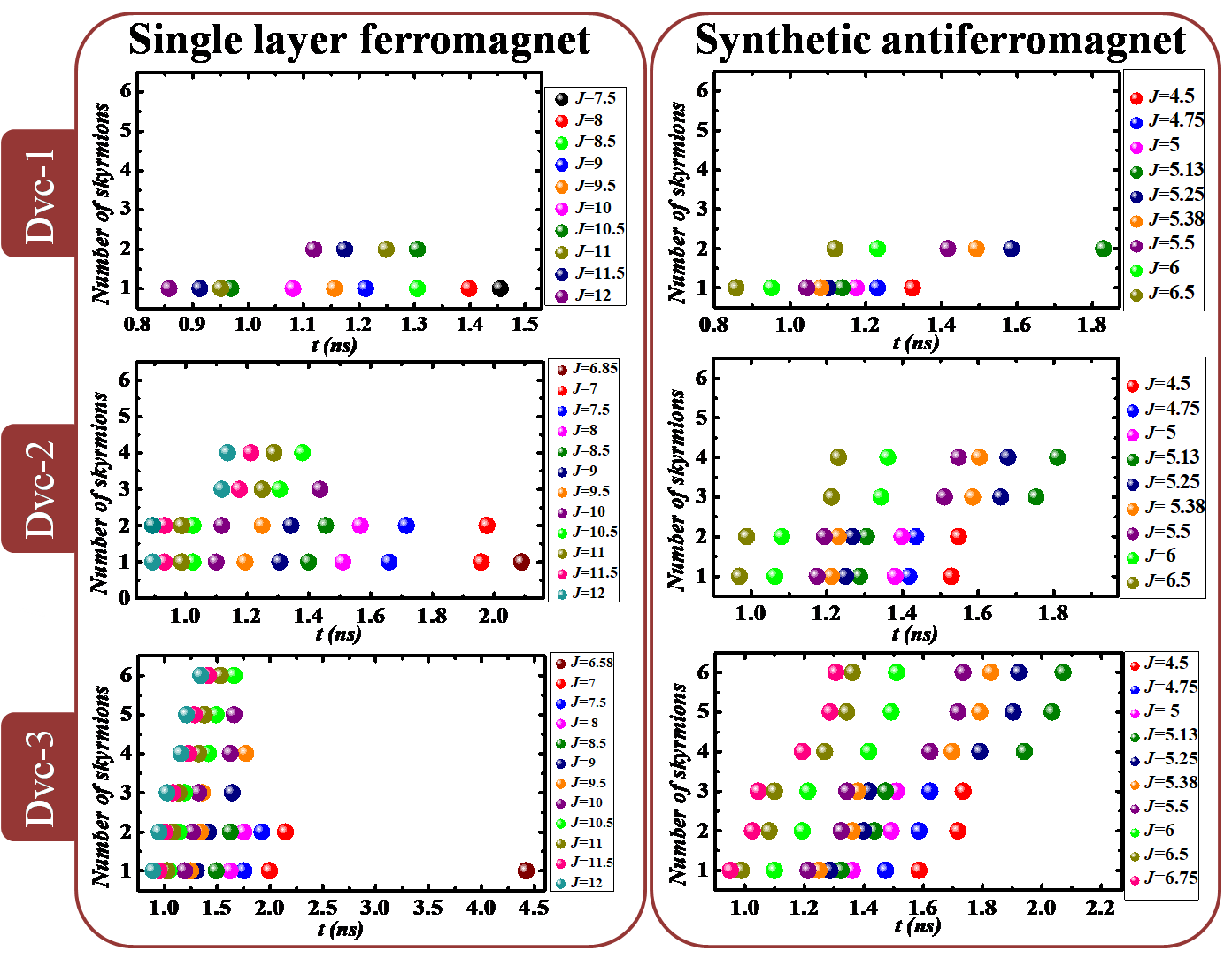}
	\renewcommand{\thefigure}{S\arabic{figure}}
	\caption{ Number of skyrmions as a function of time ($t$) for maximum states achieved per device is plotted for single FM and SAF at different values of $J (\times 10^{12} A/m^2$).}
\end{figure}

\leavevmode\thispagestyle{empty}\newpage

The increase in topological number as function of time (in ns) corresponding to creation of skyrmion is achieved in (a) SAF and (b) single FM layer. Fig.S4 (b) shows transition of topological number from Q=0 to higher skyrmionic states as the value of $J$ is increased in three different devices in single FM. [The conversion processes at different value of $J$ are shown in gifs where formation of all possible number of skyrmions can be observed.]

\begin{figure}[h!]
	\centering
	\includegraphics[width=0.9\linewidth]{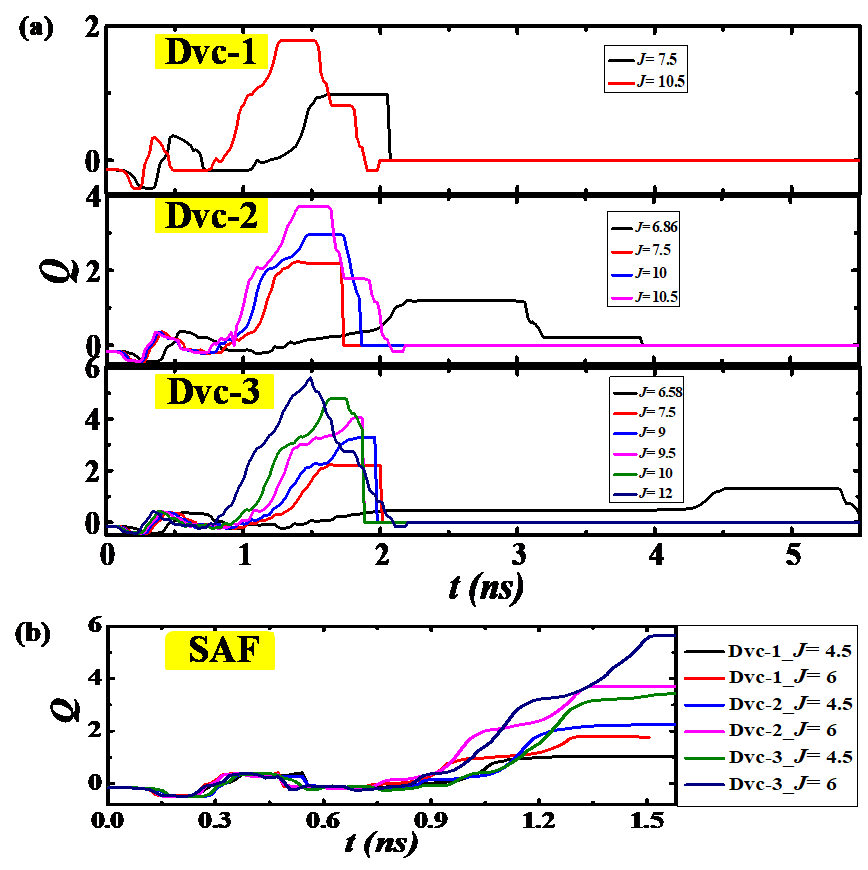}
	\renewcommand{\thefigure}{S\arabic{figure}}
	\caption{ The variation of total topological number as a function of time per device is plotted for (a) single FM layer and (b) SAF at different value of  $J (\times 10^{12} A/m^2$).}
\end{figure}

\leavevmode\thispagestyle{empty}\newpage

The exchange energy (Fig. S5 (a)) and PMA energy (Fig. S5 (b)) show a similar trend throughout conversion for different devices.  The $E_{ex}$ and $E_{PMA}$ decrease with the reduction of total length of the magnetic DWs until skyrmionium transforms to DWs. The DW length of skyrmionium is longer as compared to a skyrmion. After conversion, as the number of skyrmions increase in the system, the exchange and PMA energies also increases. However, the $E_{DMI}$ shows an inverse trend as compared to $E_{ex}$ and $E_{PMA}$, represented in fig. S5 (c). The increase in exchange energy and lowering of DMI energy on generation of more skyrmionic states can be explained by the fact that the exchange energy favors collinear spin configuration, whereas, the DMI prefer non-collinear spin structure. In fig. S5 (d), initially the $E_{demag}$ increases during the transformation of skyrmionium to DWs. But, as the number of skyrmions in the system increase, the demagnetization energy decreases. In case of Dvc-3, the demagnetization energy becomes almost equivalent to the initial value. However, the larger FM region of Dvc-2, as compared to Dvc-1 and Dvc-3, corresponds to higher demagnetization energy. Hence, the total micromagnetic energy of Dvc-2 is higher than Dvc-1 and Dvc-3. Fig. S5 (e) shows the graph for number of DW spins as a function of time, where the number of DW spins corresponds to the total length of the DW in the system. This matches with the exchange energy and PMA energy profiles. During conversion, the graphs initially decrease when there is reduction of total length of the magnetic domain walls forming the topological objects. But, in Dvc-3, the first skyrmion starts to annihilate at the device end as the last skyrmion is released. Hence, the lengths of DW reduce leading to slightly lower count of DW spins than expected.

\begin{figure}[h!]
	\centering
	\includegraphics[width=0.9\linewidth]{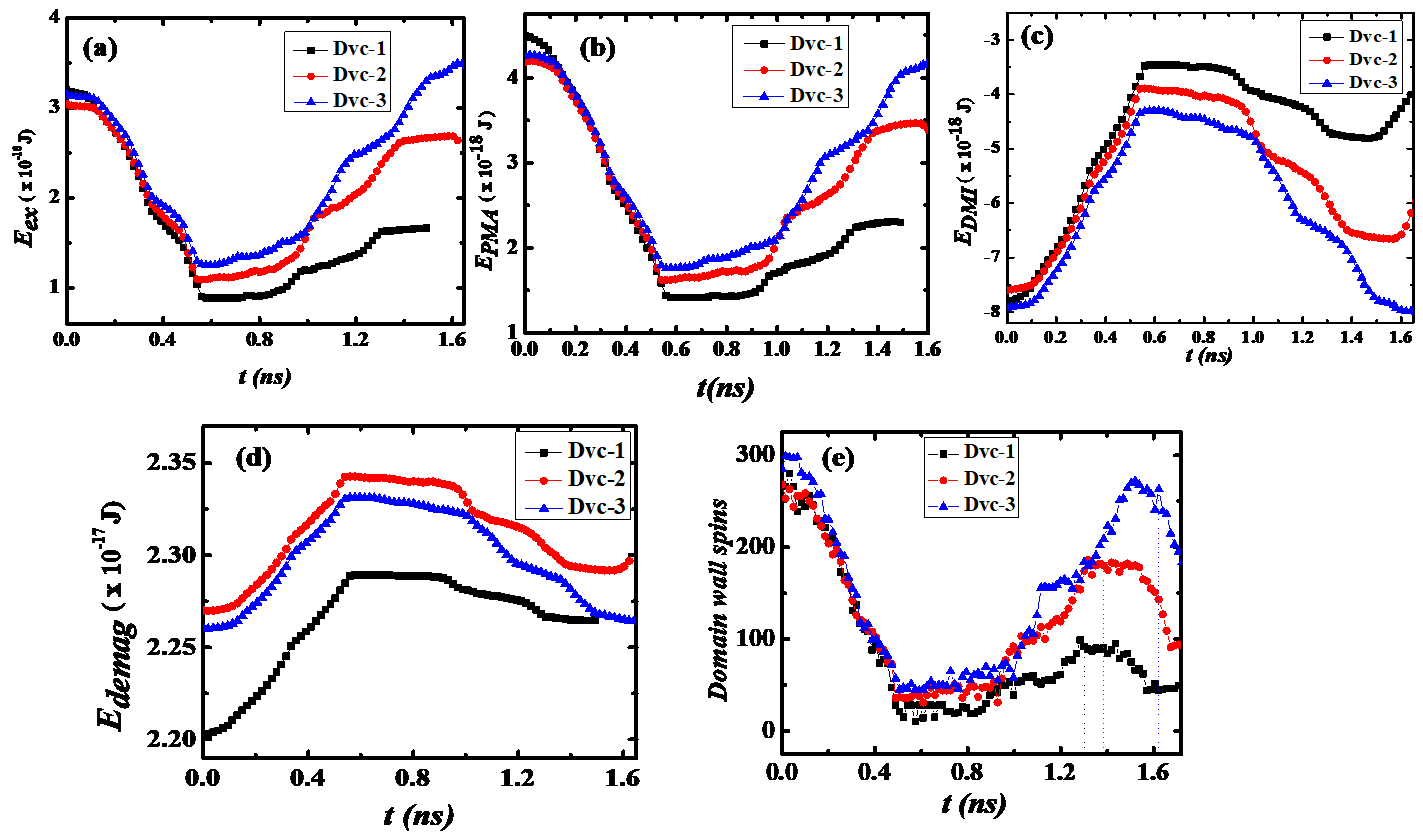}
	\renewcommand{\thefigure}{S\arabic{figure}}
	\caption{(a), (b), (c), (d) and (e) show the exchange energy, the PMA energy, DMI energy, demagnetization energy and domain wall spins as a function of $t$, respectively at $J = 10.5 \times 10^{12} A/m^2$ for single FM layer.}
\end{figure}


\bibliography{references-supp}